\documentclass[lettersize,hidelinks,journal]{IEEEtran}

\usepackage{cite}
\usepackage{amsmath,amssymb,amsfonts}
\usepackage{graphicx}
\usepackage{textcomp}
\usepackage{xcolor}
\usepackage[acronym,shortcuts]{glossaries}
\usepackage{bm}
\usepackage{caption}
\usepackage{subfigure}
\usepackage{relsize}
\usepackage{soul}
\usepackage{algorithm, algpseudocode}
\usepackage{algcompatible}
\usepackage{outlines}
\usepackage{orcidlink}
\usepackage{lipsum,comment}
\usepackage{mathtools}
\usepackage{tabularx,flushend}
\def\BibTeX{{\rm B\kern-.05em{\sc i\kern-.025em b}\kern-.08em
T\kern-.1667em\lower.7ex\hbox{E}\kern-.125emX}}

\newcommand{\trans}[0]{^{\mathsf{T}}}

\newcommand{\herm}[0]{^{\mathsf{H}}}

\newacronym{RPE}{RPE}{radar parameter estimation}
\newacronym{OTFS}{OTFS}{orthogonal time frequency space}
\newacronym{AFDM}{AFDM}{affine frequency division multiplexing}
\newacronym{MIMO}{MIMO}{multiple-input multiple-output}
\newacronym{SISO}{SISO}{single-input single-output}
\newacronym{ISAC}{ISAC}{integrated sensing and communications}
\newacronym{3D}{3D}{three-dimensional}
\newacronym{2D}{2D}{two-dimensional}
\newacronym{1D}{1D}{one-dimensional}
\newacronym{RX}{RX}{receiver}
\newacronym{TX}{TX}{transmitter}
\newacronym{BF}{BF}{beamforming}
\newacronym{mmWave}{mmWave}{millimeter-wave}
\newacronym{SotA}{SotA}{state-of-the-art}
\newacronym{ULA}{ULA}{uniform linear array}
\newacronym{QAM}{QAM}{quadrature amplitude modulation}
\newacronym{ISFFT}{ISFFT}{inverse symplectic finite Fourier transform}
\newacronym{SFFT}{SFFT}{symplectic finite Fourier transform}
\newacronym{AWGN}{AWGN}{additive white Gaussian noise}
\newacronym{OFDM}{OFDM}{orthogonal frequency division multiplexing}
\newacronym{OCDM}{OCDM}{orthogonal chirp division multiplexing}
\newacronym{BS}{BS}{base station}
\newacronym{UE}{UE}{user equipment}
\newacronym{DFT}{DFT}{discrete Fourier transform}
\newacronym{IDFT}{IDFT}{inverse discrete Fourier transform}
\newacronym{TD}{TD}{time-domain}
\newacronym{wlg}{wlg}{without loss of generality}
\newacronym{CP}{CP}{cyclic prefix}
\newacronym{DAFT}{DAFT}{discrete affine Fourier transform}
\newacronym{IDAFT}{IDAFT}{inverse discrete affine Fourier transform}
\newacronym{CPP}{CPP}{\textit{chirp-periodic} prefix}
\newacronym{IDZT}{IDZT}{inverse discrete Zak transform}
\newacronym{DZT}{DZT}{discrete Zak transform}
\newacronym{ICI}{ICI}{inter-carrier interference}
\newacronym{BER}{BER}{bit error rate}
\newacronym{DoF}{DoF}{degrees-of-freedom}
\newacronym{FD}{FD}{full-duplex}
\newacronym{SIMO}{SIMO}{single-input multiple-output}
\newacronym{MISO}{MISO}{multiple-input single-output}
\newacronym{AoD}{AoD}{angle-of-departure}
\newacronym{AoA}{AoA}{angle-of-arrival}
\newacronym{RF}{RF}{radio frequency}
\newacronym{SIM}{SIM}{stacked intelligent metasurfaces}
\newacronym{FIM}{FIM}{flexible intelligent metasurface}
\newacronym{FPGA}{FPGA}{field programmable gate array}
\newacronym{UPA}{UPA}{uniform planar array}
\newacronym{CC}{CC}{communication-centric}
\newacronym{I/O}{I/O}{input-output}
\newacronym{iid}{i.i.d.}{independent and identically distributed}
\newacronym{IoT}{IoT}{internet of things}
\newacronym{V2X}{V2X}{vehicle-to-everything}
\newacronym{NTN}{NTN}{non-terrestrial network}
\newacronym{LEO}{LEO}{low-earth orbit}
\newacronym{THz}{THz}{terahertz}
\newacronym{EM}{EM}{electromagnetic}
\newacronym{RIS}{RIS}{reconfigurable intelligent surface}
\newacronym{DoA}{DoA}{direction-of-arrival}
\newacronym{DD}{DD}{doubly-dispersive}
\newacronym{ODDM}{ODDM}{orthogonal delay-Doppler division multiplexing}
\newacronym{LoS}{LoS}{line-of-sight}
\newacronym{NLoS}{NLoS}{non-line-of-sight}
\newacronym{6G}{6G}{sixth generation}
\newacronym{MPDD}{MPDD}{metasurfaces-parameterized DD}
\newacronym{FPDD}{FPDD}{FIM-parameterized DD}
\newacronym{GaBP}{GaBP}{Gaussian Belief Propagation}
\newacronym{MSE}{MSE}{mean-squared-error}
\newacronym{sIC}{soft IC}{soft interference cancellation}
\newacronym{soft RG}{soft RG}{soft replica generation}
\newacronym{BG}{BG}{belief generation}
\newacronym{SGA}{SGA}{scalar Gaussian approximation}
\newacronym{CLT}{CLT}{central limit theorem}
\newacronym{PDF}{PDF}{probability density function}
\newacronym{QPSK}{QPSK}{quadrature phase-shift keying}
\newacronym{LMMSE}{LMMSE}{linear minimum mean square error}
\newacronym{SNR}{SNR}{signal-to-noise ratio}
\newacronym{QoS}{QoS}{quality of service}

\newacronym{B5G}{B5G}{beyond fifth generation}
\newacronym{VR}{VR}{virtual reality}
\newacronym{XR}{XR}{extended reality}
\newacronym{ITN}{ITN}{intelligent traffic networks}
\newacronym{SAGIN}{SAGIN}{space-air-ground integrated network}
\newacronym{UAV}{UAV}{unmanned aerial vehicle}
\newacronym{MUSIC}{MUSIC}{Multiple Signal Classification}
\newacronym{ICC}{ICC}{integrated communication and computing}

\begin{document}

\title{Flexible Intelligent Metasurfaces in High-Mobility MIMO Integrated Sensing and Communications}

\author{Kuranage Roche Rayan Ranasinghe\textsuperscript{\orcidlink{0000-0002-6834-8877}},~\IEEEmembership{Graduate Student Member,~IEEE,}
Jiancheng An\textsuperscript{\orcidlink{0000-0003-2512-9989}},~\IEEEmembership{Member,~IEEE,} \\
Iv{\'a}n Alexander Morales Sandoval\textsuperscript{\orcidlink{0000-0002-8601-5451}},~\IEEEmembership{Graduate Student Member,~IEEE,}
Hyeon Seok Rou\textsuperscript{\orcidlink{0000-0003-3483-7629}},~\IEEEmembership{Member,~IEEE,} \\
Giuseppe Thadeu Freitas de Abreu\textsuperscript{\orcidlink{0000-0002-5018-8174}},~\IEEEmembership{Senior Member,~IEEE,} Chau Yuen\textsuperscript{\orcidlink{0000-0002-9307-2120}},~\IEEEmembership{Fellow,~IEEE,} \\
and M\'{e}rouane Debbah\textsuperscript{\orcidlink{0000-0001-8941-8080}},~\IEEEmembership{Fellow,~IEEE}
% <-this % stops a space
\thanks{K. R. R. Ranasinghe, I. A. M. Sandoval, H. S. Rou and G. T. F. de Abreu are with the School of Computer Science and Engineering, Constructor University (previously Jacobs University Bremen), Campus Ring 1, 28759 Bremen, Germany (emails: \{kranasinghe,imorales,hrou,gabreu\}@constructor.university).} 
\thanks{J. An and C. Yuen is with the School of Electrical and Electronics Engineering, Nanyang Technological University, Singapore 639798 (e-mails:\{jiancheng.an,chau.yuen\}@ntu.edu.sg).}
\thanks{M. Debbah is with the KU 6G Research Center, Department of Computer and Information Engineering, Khalifa University, Abu Dhabi, United Arab Emirates, and also with the CentraleSup\'{e}lec, University Paris-Saclay, 91192 Gif-sur-Yvette, France (e-mail: merouane.debbah@ku.ac.ae).}\vspace{-3ex}}

%\markboth{To be submitted to the IEEE for possible publication}{Shell \MakeLowercase{\textit{et al.}}: A Sample Article Using IEEEtran.cls for IEEE Journals}


\maketitle

\begin{abstract}
We propose a novel \ac{DD} \ac{MIMO} channel model incorporating \acp{FIM}, which is suitable for \ac{ISAC} in high-mobility scenarios. 
We then discuss how the proposed \ac{FPDD} channel model can be applied in a logical manner to \ac{ISAC} waveforms that are known to perform well in \ac{DD} environments, namely, \ac{OFDM}, \ac{OTFS}, and \ac{AFDM}. 
Leveraging the proposed model, we formulate an achievable rate maximization problem with a strong sensing constraint for all the aforementioned waveforms, which we then solve via a gradient ascent algorithm with closed-form gradients presented as a bonus.
Our numerical results indicate that the achievable rate is significantly impacted by the emerging \ac{FIM} technology with careful parametrization essential in obtaining strong \ac{ISAC} performance across all waveforms suitable to mitigating the effects of DD channels.
\end{abstract}


\glsresetall

\section{Introduction}
%% Intro. to doubly-dispersive channels (high mobility and CC ISAC)
%% Into to 6G , ....
\IEEEPARstart{N}{ext}-generation wireless networks in the \ac{B5G} and \ac{6G} are expected to meet unprecedented performance demands and support a diverse range of applications, such as \ac{ISAC} \cite{LuongCOMMST2025}, \ac{ICC} \cite{RanasingheICNC2025_comp,ranasinghe2025flexibledesignframeworkintegrated}, \ac{VR}/\ac{XR} \cite{YuNWTWORK2023}, \ac{ITN} \cite{NguyenJSAC2024}, \ac{SAGIN} \cite{CuiCC2022}, holographic communications \cite{DengJSAC2023}, and massive \ac{IoT} \cite{Chowdhury_6G,Dang_6G}.

As expected above, \ac{6G} networks will increasingly operate in challenging wireless channels, i.e., in high-mobility multipath applications such as \ac{V2X}, high-speed rail, \ac{UAV}, \ac{LEO} satellite, and \ac{SAGIN} links which involve nodes moving at significant velocities \cite{GiordaniCOMMAG2020}.
Such environments are well-known to exhibit channel responses that are both time-selective and frequency-selective (often described as time-frequency selective fading, or doubly-dispersive channels), which incorporate both the temporal variation (Doppler shifts) and frequency dispersion (multipath delays) \cite{Bliss_Govindasamy_2013}.
Therefore, modeling beyond the quasi-static or frequency-flat assumptions commonly used in conventional wireless communication system models are essential to accurately capture the propagation effects in these scenarios \cite{TariqWCOMM2020,Rou_SPM_2024}. 

Notably, the inherent delay-Doppler structure of such channels can be leveraged for communication robustness and for \ac{ISAC} \cite{Hadani_WCNC_2017, liu2022integrated} - for example, enabling radar parameter estimation (e.g., target range and velocity) directly from communication waveforms \cite{NguyenJSTSP2024,XiaoTSP2024, RanasingheARXIV2024, LuoIoTJ2025}.
This opens the door to native and efficient \emph{communication-centric} \ac{ISAC} techniques that unify data transmission and environmental sensing within the same infrastructure \cite{cheng2022integrated}, highlighting the need for the development of models capable of leveraging these properties.

%% EM strcutures

On the other hand, instead of channel modeling, another very promising approach to achieve robustness and control over the challenging wireless channel is to directly manipulate the propagation environment itself using reconfigurable electromagnetic structures, which has been attracting great attention in the recent years \cite{AlexandropolousVTM2024, StutzOJCOMS2025, dardari2025overtheairmultifunctionalwidebandelectromagnetic, AtaloglouTAP2025}.
The most standard and widely-investigated example is the \ac{RIS} -- which are planar metasurfaces with tunable meta-atom elements -- having been extensively studied as a means to engineer reflections and enhance wireless coverage and capacity in an energy-efficient manner \cite{LSA_2014_Cui_Coding, KolomvakisTWC2025, DrouliasTWC2024, ZhangTWC2025}. 
More recently, enhanced metasurface architectures such as \ac{SIM} have been developed to perform elaborate signal processing directly in the electromagnetic wave domain \cite{AnWC2024, AXN2023}, which comprises a large number of tunable meta-atoms collected into layers and \emph{stacked} very close to one another.

%% FIM: Added by Jiancheng, please check the conherence with other parts

Building upon these advances, an emerging paradigm is the \emph{flexible} variant, known as the \ac{FIM}, a metasurface that can physically reconfigure or morph its \ac{3D} surface shape in response to time-varying wireless channel conditions \cite{TWC_2024_An_Flexible, Nature_2022_Bai_A, TAP_2025_An_Emerging, Sci_2015_Ni_An}. 
Unlike traditional \acp{RIS} that rely solely on electronically controlled phase shifts, \acp{FIM} can adjust the positions of their meta-atoms in the normal direction of the planar surface. 
This synergistic combination of physical and electronic flexibility provides \acp{FIM} with enhanced spatiotemporal control capabilities that exceed those of conventional \acp{RIS} with static structures \cite{NC_2016_Kamali_Decoupling}. 
For instance, Ni \emph{et al.} \cite{NC_2022_Ni_Soft} have developed a reconfigurable \ac{FIM} that uses  a liquid metal soft microfluidic network embedded within an elastomeric matrix, controlled by electromagnetic actuation. 
This innovative design allows for real-time and programmable surface shape morphing with fully reversible properties. 
More recently, another innovative \ac{FIM} was constructed using an array of tiny metallic filaments \cite{Nature_2022_Bai_A}, actuated by reprogrammable distributed Lorentz forces generated by electrical currents passing through a static magnetic field. 
This setup grants the \ac{FIM} precise and rapid dynamic morphing capabilities, enabling it to promptly altering its structure\footnote{A video showing the real-time morphing ability of an \ac{FIM} can be found at \url{https://www.eurekalert.org/multimedia/950133}.}. 
Additionally, incorporating a mechanical locking mechanism can sustain the newly morphed shapes.

Recent groundbreaking research has unveiled the considerable benefits of \ac{FIM} technology in wireless communication and sensing systems \cite{TCOM_2025_An_Flexible, TVT_2025_Teng_Flexible}. 
In \cite{TCOM_2025_An_Flexible}, An \emph{et al.} explored the function of \acp{FIM} as \ac{MIMO} transceiver arrays. 
They characterized the capacity limits of \ac{FIM}-aided \ac{MIMO} transmissions over frequency-flat fading channels by jointly optimizing the \ac{3D} surface shape configurations of both transmitting and receiving \acp{FIM}, as well as the transmit covariance matrix. 
The numerical results demonstrated that \acp{FIM} can double the \ac{MIMO} capacity compared to conventional rigid arrays under specific system settings.
Furthermore, they investigated \ac{FIM}-assisted multiuser downlink communications \cite{TWC_2024_An_Flexible}, focusing on minimizing the transmit power at the \ac{BS} by jointly optimizing transmit beamforming and the \ac{FIM} surface configuration, while meeting the \ac{QoS} requirements for all users and adhering to the maximum morphing range of the \ac{FIM}. 
Indeed, simulation results showed that the \ac{FIM} can reduce the required transmit power by nearly $3$ dB compared to conventional rigid \ac{2D} arrays, while maintaining the same data rate. 
More recently, Teng \emph{et al.} \cite{TVT_2025_Teng_Flexible} examined the impact of \ac{FIM} on wireless sensing performance under the per-antenna power constraint. 
Their findings indicated that the cumulated power of the probing signals at target locations was improved by $3$ dB by optimizing the surface shape of the transmitting \ac{FIM}.

%% combining FIM with DD

However, to fully harness \acp{FIM} in high-mobility and \ac{MIMO} \ac{ISAC} contexts, it is crucial to integrate their functionality with appropriate channel models and system design frameworks.
Many existing studies on intelligent metasurfaces assume simplified channel conditions (e.g., static or block-fading channels and frequency-flat propagation), which may not hold in fast-varying environments. 
Therefore, the joint consideration of metasurface reconfigurability and doubly-dispersive channel behavior is largely uncharted, albeit essential for realizing the full potential of metasurfaces in practical scenarios.

%% SotA on combining

A recent effort by Ranasinghe \emph{et al.} \cite{ranasinghe2025metasurfacesintegrateddoublydispersivemimochannel,ranasinghe2025doublydispersivemimochannelsstacked} took a step in this direction by formulating a parametric \ac{DD} channel model that includes metasurface-based transceivers (\acp{SIM} at the transmitter and receiver) and intermediate \ac{RIS} panels in the environment, which demonstrated how metasurface parameters can be incorporated into the delay-Doppler domain channel characterization, allowing the design of waveforms (such as \ac{OFDM}, \ac{OTFS} and \ac{AFDM}) and transceiver strategies tailored to time-varying multipath conditions, with a corresponding \ac{ISAC} analysis done in \cite{ranasinghe2025parametrizedstackedintelligentmetasurfaces}.

However, the paradigm of flexible metasurfaces introduces additional challenges and opportunities that remain open. 
In particular, the physical shape-shifting capability of \acp{FIM} add a new layer of complexity to channel modeling and optimization, as the channel response itself can be altered in real time by the metasurface geometry.
Therefore in this article, we explore the integration of \ac{FIM} technology with doubly-dispersive channel modeling for the \ac{B5G}/\ac{6G} \ac{MIMO} \ac{ISAC} paradigm. 

%% Contributions
The contributions of this paper are summarized as follows:
%% Contributions
\begin{itemize}
\item We describe a novel \ac{FPDD} channel model which can be applied to the design of \ac{ISAC} waveforms known to perform well in \ac{DD} environments, such as \ac{OFDM}, \ac{OTFS}, and \ac{AFDM}, as well as to derive their complete \ac{I/O} relationships in the corresponding domains.
\item We formulate and solve -- via a gradient ascent algorithm with corresponding closed-form gradients -- an achievable rate maximization problem with a strong sensing constraint for all the aforementioned waveforms and under the given model. We highlight that both the model and the optimization problem are contributions on their own.
\item Simulation results indicate that the achievable rate is significantly impacted by the emerging \ac{FIM} technology with careful parametrization essential in obtaining strong \ac{ISAC} performance across all waveforms suitable to mitigating the effects of DD channels.
\end{itemize}

\textit{Notation:} All scalars are represented by upper or lowercase letters, while column vectors and matrices are denoted by bold lowercase and uppercase letters, respectively.
The diagonal matrix constructed from vector $\mathbf{a}$ is denoted by diag($\mathbf{a}$), while $\mathbf{A}\trans$, $\mathbf{A}\herm$, $\mathbf{A}^{1/2}$, and $[\mathbf{A}]_{i,j}$ denote the transpose, Hermitian, square root and the $(i,j)$-th element of a matrix $\mathbf{A}$, respectively.
The convolution and Kronecker product are respectively denoted by $*$ and $\otimes$, while $\mathbf{I}_N$ and $\mathbf{F}_N$ represent the $N\times N$ identity and the normalized $N$-point \ac{DFT} matrices, respectively.
The sinc function is expressed as $\text{sinc}(a) \triangleq \frac{\sin(\pi a)}{\pi a}$, and $\jmath\triangleq\sqrt{-1}$ denotes the elementary complex number.

%%%%%%%%%%%%%%%%%%%%%%%%%%%%%%%%%%%%%%%%%%%%%%%%%%%%%
\section{The Proposed FIM MIMO Channel Model}
\label{FIM_MIMO_Model}

\begin{figure*}[t!]
\centering
\includegraphics[width=0.9\textwidth]{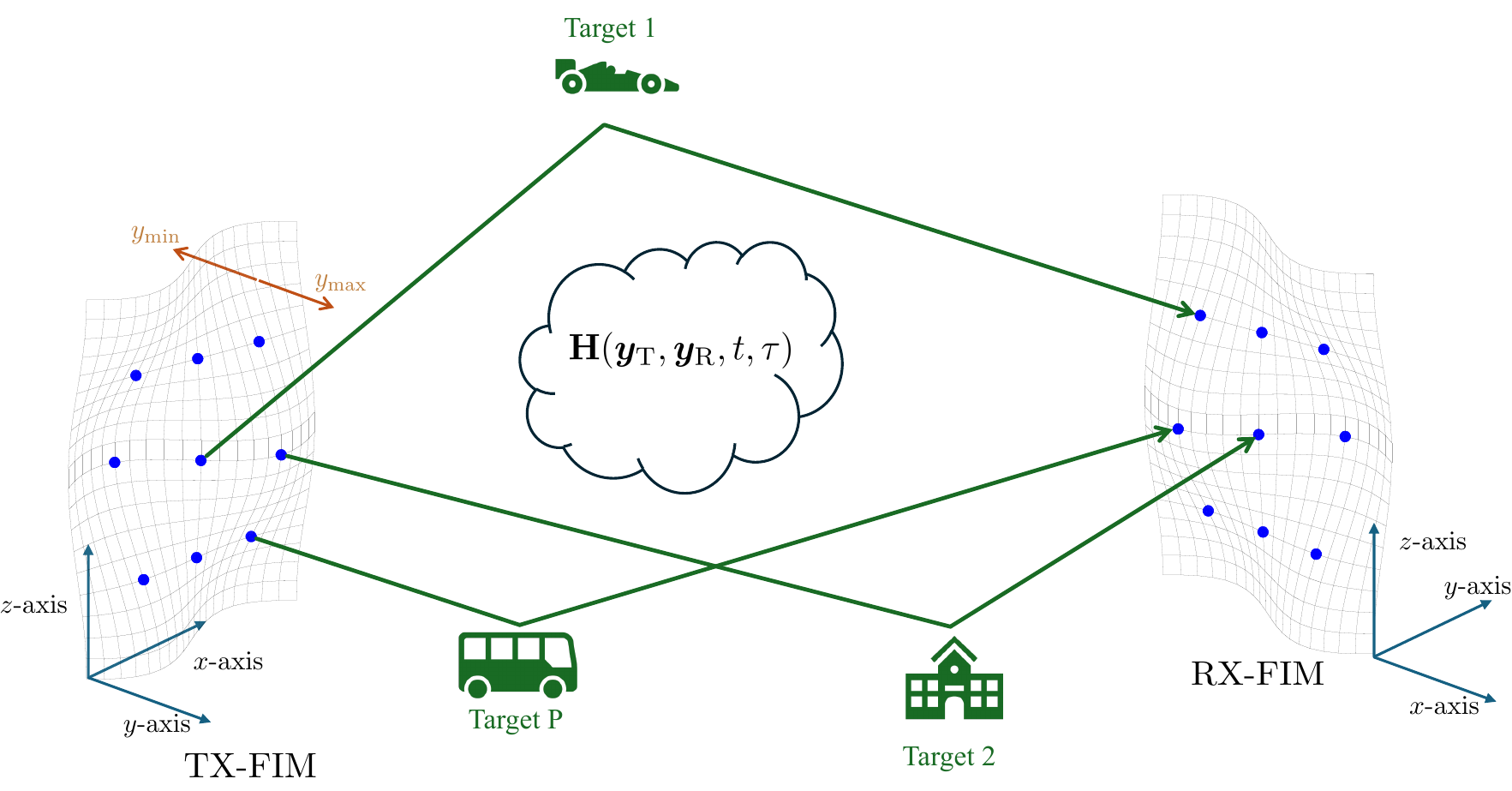}
\vspace{-1ex}
\caption{The considered \ac{FIM} \ac{MIMO} system for high-mobility scenarios, which includes two \acp{FIM}, one acting as the \ac{TX} and the other as the \ac{RX}, with 3 scatterers of interest.}
\label{fig:system_model_FIM}
\vspace{-2ex}
\end{figure*}

\subsection{FIM Antenna Array Response}

Let $\theta\in [0,\pi]$ and $\phi \in [-\frac{\pi}{2},\frac{\pi}{2}]$ denote the arbitrary \ac{AoA} or \ac{AoD} elevation and azimuth angles~\cite{AnJSAC2023,AnTWC2025} of a channel propagation path to (or from) a flexible \ac{UPA} with $B \triangleq B_x B_z$ elements\footnote{Without loss of generality, the \ac{UPA} is aligned parallel to the $y$ direction with elements occupying space in the $x$ and $z$ dimensions. The generalization to arbitrary axes is trivial; some other orientations are discussed in \cite{AnJSAC2024}.}, where $B_x$ and $B_z$ refer to the number of antenna elements along the $x$-axis and $y$-axis, respectively, with $\mathcal{B} \triangleq \{1,2,\dots,B\}$ denoting the set of antenna elements.

The key principle of a \ac{FIM} stems from the fact that each radiating element is allowed some movement in an additional third dimension; $i.e.,$ the $y$-axis via a connected controller.
Let $\bm{p}_b = [x_b, y_b, z_b]\trans \in \mathbb{R}^3, \forall b \in \mathcal{B}$ be the tuple denoting the location of the $b$-th radiating element, where each element is modelled as a point \cite{AnTWC2025}.
Using the primary element as the reference leads to
\begin{subequations}
\begin{equation}
        \label{eq:x_def_elem}
        x_b = d_x \times \text{mod}(b - 1,B_x), \forall b \in \mathcal{B},
\end{equation}
\begin{equation}
        \label{eq:z_def_elem}
        z_b = d_z \times \lfloor (b - 1)/B_x) \rfloor, \forall b \in \mathcal{B},
\end{equation}
with $d_x$ and $d_z$ being the element spacing in the \ac{UPA}'s $x$- and $z$-axis directions, respectively, which are usually set as $d_x = d_z = \lambda/2$.

Additionally, the $y$-coordinate of each radiating element satisfies the constraint~\cite{AnTWC2025}
\begin{equation}
    \label{eq:y_def_constr}
    y_\text{min} \leq y_b \leq y_\text{max}, \forall b \in \mathcal{B},
\end{equation}
\end{subequations}
where $y_\text{min}$ and $y_\text{max}$ represent the minimum and maximum $y$-coordinate allowed for each radiating element, and we define $\zeta \triangleq y_\text{min} - y_\text{max} > 0$ to be the \textit{morphing range} characterizing the \ac{FIM}, where we 
henceforth set \Ac{wlg}, $y_\text{min} = 0$, \Ac{wlg}.

Subsequently, the distinguishing shape of a \ac{FIM} is given by
\begin{equation}
    \label{eq:FIM_surface_shape}
    \bm{y} \triangleq [y_1, y_2, \dots, y_B]\trans \in \mathbb{R}_+^{B \times 1}.
\end{equation}
%% DOUBLE check the domain of y here!!! It is said to be complex in the paper!!

Then, the response vector $\mathbf{b}(\bm{y},\phi,\theta) \in \mathbb{C}^{B \times 1}$ corresponding to a path impinging onto (or outgoing from) the array is given as 
\begin{align}
\mathbf{b}(\bm{y},\phi,\theta) \!&\triangleq\! \tfrac{1}{\sqrt{B}} \!\Big[\! 1, e^{\jmath \frac{2\pi}{\lambda} \big( x_1 \sin(\theta) \cos(\phi) + y_1 \sin(\theta) \sin(\phi) + z_1 \cos(\theta) \big)}\!\!,  \nonumber \\
&\hspace{-5ex} \dots, e^{\jmath \frac{2\pi}{\lambda} \big( x_B \sin(\theta) \cos(\phi) + y_B \sin(\theta) \sin(\phi) + z_B \cos(\theta) \big)} \Big]\trans.
\label{eq:steering_vec_FIM} 
\end{align}
%% CHECK if the square root exists here!!!!!!!

\subsection{FIM MIMO Channel Model}

Leveraging the above, the \ac{MIMO} \ac{DD} channel model with \acp{FIM} at both the transmitter and receiver $\mathbf{H}(\bm{y}_\mathrm{T},\bm{y}_\mathrm{R},t,\tau) \in \mathbb{C}^{N_\mathrm{R} \times N_\mathrm{T}}$ is given by
\vspace{-1ex}
\begin{eqnarray}
\label{eq:MIMO_TD_channel_general_SIM}
\mathbf{H}(\bm{y}_\mathrm{T},\bm{y}_\mathrm{R},t,\tau) \triangleq \sqrt{\tfrac{N_\mathrm{T}N_\mathrm{R}}{P}} \sum_{p=1}^P h_p e^{\jmath 2\pi \nu_p t} \delta\left(\tau-\tau_p\right)&&\\
&&\hspace{-34ex}\times  \mathbf{b}_{\mathrm{R}:p}\left(\bm{y}_\mathrm{R},\phi_p^{\rm in},\theta_p^{\rm in}\right) \mathbf{b}_{\mathrm{T}:p}\herm\left(\bm{y}_\mathrm{T},\phi_p^{\rm out},\theta_p^{\rm out}\right),\nonumber
\vspace{-1ex}
\end{eqnarray}
where $\mathbf{b}_{\mathrm{T}:p}(\cdot,\cdot,\cdot) \in \mathbb{C}^{N_\mathrm{T} \times 1}$ and $\mathbf{b}_{\mathrm{R}:p}(\cdot,\cdot,\cdot) \in \mathbb{C}^{N_\mathrm{R} \times 1}$ are respectively the \ac{FIM} \ac{UPA} response vectors for the \ac{TX} and \ac{RX} defined in equation \eqref{eq:steering_vec_FIM} containing $N_\mathrm{T}$ and $N_\mathrm{R}$ elements, with $(\bm{y}_\mathrm{R},\phi_p^{\rm in},\theta_p^{\rm in})$ and $(\bm{y}_\mathrm{T},\phi_p^{\rm out},\theta_p^{\rm out})$ containing the \ac{FIM} shaping vector and pairs of azimuth and elevation \acp{AoA} and \acp{AoD}, respectively, for each $p$-th signal propagation path having the complex channel gain $h_p$, with $p=\{1,\ldots,P\}$. 
In the above, $\tau_p \in [0,\tau_\text{max}]$ and $\nu_p \in [-\nu_\text{max},\nu_\text{max}]$ denote each $p$-th path's delay in seconds and Doppler shift in Hz, respectively\footnote{While the motion of the \acp{FIM} also causes slight variations in the relative $\tau_p$ and $\nu_p$, this effect is neglegible when the \ac{FIM} movement is restricted to be a factor of $\lambda$ as in \cite{AnTWC2025}.}.

%%%%%% CAREFUL HERE !!!!!!!!!
\begin{figure*}[t!]
\setcounter{equation}{5}
\normalsize
\begin{align}
\mathbf{r}[n]\! =\! \sum_{\ell=0}^\infty \bigg[ \bigg( \sum_{p=1}^P \underbrace{\sqrt{\tfrac{N_\mathrm{T}N_\mathrm{R}}{P}} h_p \mathbf{U}\herm  
\mathbf{b}_{\mathrm{R}:p}\left(\bm{y}_\mathrm{R},\phi_p^{\rm in},\theta_p^{\rm in}\right) \mathbf{b}_{\mathrm{T}:p}\herm\left(\bm{y}_\mathrm{T},\phi_p^{\rm out},\theta_p^{\rm out}\right)
\mathbf{V}}_{\check{\mathbf{H}}_p(\bm{y}_\mathrm{T},\bm{y}_\mathrm{R}) \in \mathbb{C}^{d_s \times d_s}}  e^{\jmath 2\pi f_p \frac{n}{N}}  \delta[ \ell\! -\! \ell_p ] \bigg)  \mathbf{s}[n\! -\! \ell] \bigg]\! + \!\mathbf{w}[n]
\label{eq:sampled_TD}
\end{align}
\hspace{30ex} \hrulefill \hspace{30ex}
\begin{align}
\label{eq:vectorized_TD_IO}
\mathbf{r}_v = \sum_{u=1}^{d_s}  \overbrace{\sum_{p=1}^P \check{h}_{p,v,u}  \underbrace{\mathbf{\Theta}_p  \mathbf{\Omega}^{f_p}  \mathbf{\Pi}^{\ell_p}}_{\triangleq\mathbf{G}_p \in \mathbb{C}^{N \times N}}}^{\triangleq\bar{\mathbf{H}}_{v,u}(\bm{y}_\mathrm{T},\bm{y}_\mathrm{R}) \in \mathbb{C}^{N \times N}} \mathbf{s}_u + \mathbf{w}_v 
= \sum_{u=1}^{d_s} \bar{\mathbf{H}}_{v,u}(\bm{y}_\mathrm{T},\bm{y}_\mathrm{R})  \mathbf{s}_u + \mathbf{w}_v
%\in \mathbb{C}^{N \times 1},
\end{align}
\hspace{30ex} \hrulefill \hspace{30ex}
\normalsize
\begin{equation}
\label{eq:diagonal_CP_matrix_def}
\mathbf{\Theta}_p \triangleq \text{diag}\Big( [ \underbrace{e^{-\jmath 2\pi {\phi_\mathrm{CP}(\ell_p)}}, e^{-\jmath 2\pi {\phi_\mathrm{CP}(\ell_p - 1)}}, \dots, e^{-\jmath 2\pi {\phi_\mathrm{CP}(2)}}, e^{-\jmath 2\pi {\phi_\mathrm{CP}(1)}}}_{\ell_p \; \text{terms}}, \underbrace{1, 1, \dots, 1, 1}_{N - \ell_p \; \text{ones}}] \Big) \in \mathbb{C}^{N \times N}
\end{equation}
\hspace{30ex} \hrulefill \hspace{30ex}
\begin{equation}
\label{eq:diagonal_Doppler_matrix_def}
\boldsymbol{\Omega} \triangleq \text{diag}\Big([1,e^{-\jmath 2\pi /N},\dots,e^{-\jmath 2\pi (N-2) /N}, e^{-\jmath 2\pi (N-1) /N}]\Big) \in \mathbb{C}^{N \times N}
\end{equation}

\setcounter{equation}{4}
\hrulefill
\vspace{-3ex}
\end{figure*}

\renewcommand{\arraystretch}{1.25}
\setlength{\tabcolsep}{1.2pt}
\begin{table}[H]
\vspace{-1ex}
\caption{Variable Notation and Descriptions.}
\centering
\begin{tabular}{|c|c|}
\hline
\textbf{Variable} & \textbf{Description} \\
\hline
$N_\mathrm{T}$, $N_\mathrm{R}$ & Number of \ac{TX} and \ac{RX} \ac{FIM} elements \\
\hline
$\theta_p^{\rm in}$,$\theta_p^{\rm out}$  & \ac{AoA} and \ac{AoD} elevation angle \\
\hline
$\phi_p^{\rm in}$,$\phi_p^{\rm out}$ & \ac{AoA} and \ac{AoD} azimuth angle \\
\hline
$h_p$ & $p$-th complex channel gain \\
\hline
$\tau_p$ & $p$-th associated delay \\
\hline
$\nu_p$ & $p$-th associated Doppler shift \\
\hline
$\mathbf{b}_{\mathrm{T}:p}(\cdot,\cdot,\cdot)$ & $p$-th \ac{TX} \ac{FIM} \ac{UPA} response vector \\
\hline
$\mathbf{b}_{\mathrm{R}:p}(\cdot,\cdot,\cdot)$ & $p$-th \ac{RX} \ac{FIM} \ac{UPA} response vector \\
\hline
$\mathbf{H}(\bm{y}_\mathrm{T},\bm{y}_\mathrm{R},t,\tau)$ & \ac{FIM} \ac{MIMO} channel model \\
\hline
\end{tabular}
\label{tab:example}
\vspace{-2ex}
\end{table}

% \vspace{-1ex}
\section{I/O Relationships for OFDM, OTFS, and AFDM}
\label{IO_Model}

In this section, leveraging the previously proposed \ac{FPDD} channel model, we present various expressions for the \ac{TD} received signal corresponding to waveforms such as \ac{OFDM}, \ac{OTFS}, and \ac{AFDM}.
The framework presented is general and also incorporates hybrid analog and digital beamformers, which can be optimized for further performance improvements \cite{SrivastavaTWC2022}.

\vspace{-2ex}
\subsection{Arbitrarily Modulated Signals}
\vspace{-1ex}

The point-to-point \ac{MIMO} configuration illustrated in Fig.~\ref{fig:system_model_FIM} employs hybrid analog-digital beamformers at its $N_\mathrm{T}$-element \ac{TX} and $N_\mathrm{R}$-element \ac{RX}, following a similar structure to \cite{YanCommL2023}.

Let $\mathbf{V} \in \mathbb{C}^{N_\mathrm{T} \times d_s}$ and $\mathbf{U} \in \mathbb{C}^{N_\mathrm{R} \times d_s}$ denote the hybrid beamformers for transmission and reception, respectively, where $d_s \triangleq \text{min}(N_\mathrm{T},N_\mathrm{R})$ represents the number of independent data streams transmitted per coherent channel block. 
In the following, the complex-valued vector $\mathbf{s}(t) \in \mathbb{C}^{d_s \times 1}$ is the power-constrained transmit signal, which can employ either \ac{OFDM}, \ac{OTFS}, or \ac{AFDM} in the \ac{TD}. 

The $d_s$-element baseband signal received at time $t$ (post-hybrid combining) through the \ac{FIM} \ac{MIMO} channel is mathematically expressed as 
\vspace{-1ex}
\begin{align}
\mathbf{r}(t) 
&\triangleq  \mathbf{U}\herm \mathbf{H}(\bm{y}_\mathrm{T},\bm{y}_\mathrm{R},t,\tau)  * \mathbf{V} \mathbf{s}(t) + \mathbf{w}(t)  \nonumber
\\&= \int\limits_{-\infty}^\infty \mathbf{U}\herm \mathbf{H}(\bm{y}_\mathrm{T},\bm{y}_\mathrm{R},t,\tau) \mathbf{V} \mathbf{s}(t - \tau) d\tau + \mathbf{w}(t),
\label{eq:TD_I/O_relationship}
\vspace{-1ex}
\end{align}
where $\mathbf{w}(t) \triangleq \mathbf{U}\herm \mathbf{n}(t) \in \mathbb{C}^{d_s \times 1}$ and $\mathbf{n}(t) \in \mathbb{C}^{N_\mathrm{R} \times 1}$ represents the \ac{AWGN} vector at the \ac{RX}, with elements that are uncorrelated both spatially and temporally, each having zero mean and variance $\sigma_n^2$. 

Let $\mathbf{r}[n] \in \mathbb{C}^{d_s \times 1}$ and $\mathbf{s}[n] \in \mathbb{C}^{d_s \times 1}$, with $n \in \{ 0,\dots,N-1 \}$, be the discrete sequences derived by sampling $\mathbf{r}(t)$ and $\mathbf{s}(t)$, respectively, at a sufficiently high rate $F_S \triangleq \frac{1}{T_S}$ in Hz within a bandwidth $B$ \cite{RanasingheICNC2025_oversampling}. 
The discrete-time form of the received signal in equation \eqref{eq:TD_I/O_relationship} is presented (at the top of the page) in equation \eqref{eq:sampled_TD}, where $\ell$ denotes the normalized discrete delay index, and $f_p \triangleq \frac{N\nu_p}{F_s}$ and $\ell_p \triangleq \frac{\tau_p}{T_s}$ represent the normalized Doppler shift and the corresponding normalized discrete delay index for each $p$-th propagation path between the \ac{TX}-\ac{FIM} and \ac{RX}-\ac{FIM}, respectively.

Considering a \ac{CP} of length $N_\mathrm{CP}$ and employing circular convolution, the $N$-element discrete-time received signal in equation \eqref{eq:sampled_TD} can be reformulated \cite{Rou_SPM_2024} as shown in equation \eqref{eq:vectorized_TD_IO}, where, for simplicity, the discrete-time index is omitted (and will be henceforth). 
In this equation, the scalars $\check{h}_{p,v,u}$ with $(v,u) = \{ 1,\dots,d_s \}$ are the $(v,u)$-th elements of the matrix $\check{\mathbf{H}}_p(\bm{y}_\mathrm{T},\bm{y}_\mathrm{R})$, implicitly defined in equation \eqref{eq:sampled_TD}.
Furthermore, $\mathbf{s}_u \triangleq [s_u[0],\ldots,s_u[N-1]] \in \mathbb{C}^{N \times 1}$ and $\mathbf{w}_v \triangleq [w_v[0],\ldots,w_v[N-1]] \in \mathbb{C}^{N \times 1}$ represent the transmit signal and \ac{AWGN} vectors for the $u$-th and $v$-th streams, respectively.

Each diagonal matrix $\mathbf{\Theta}_p \in \mathbb{C}^{N \times N}$, defined in equation \eqref{eq:diagonal_CP_matrix_def}, accounts for the effect of the \ac{CP} on the $p$-th channel path, with $\phi_\mathrm{CP}(n)$ being a phase function dependent on the sample index $n \in \{ 0,\ldots,N-1 \}$, varying with the specific waveform employed. 
Additionally, the diagonal matrix $\boldsymbol{\Omega} \in \mathbb{C}^{N \times N}$, defined in equation \eqref{eq:diagonal_Doppler_matrix_def}, comprises $N$ complex roots of unity, while $\mathbf{\Pi} \in \{0,1\}^{N \times N}$ is the forward cyclic shift matrix with elements defined as
\vspace{-1ex}
\setcounter{equation}{9}
\begin{equation}
\label{eq:PiMatrix}
\pi_{i,j} \triangleq \delta_{i,j+1} + \delta_{i,j-(N-1)}\,\;\; \delta _{ij} \triangleq
\begin{cases}
0 & \text{if }i\neq j\\
1 & \text{if }i=j
\end{cases}.
\vspace{-1ex}
\end{equation}

Using the Kronecker product to combine all $d_s$ $\mathbf{r}_v$ vectors in equation \eqref{eq:vectorized_TD_IO}, the resulting $N d_s$-element vector for the overall received signal in the \ac{TD}, for an arbitrarily modulated transmit signal, is expressed as
\vspace{-1ex}
\begin{equation}
\mathbf{r}_\mathrm{TD} = \bar{\mathbf{H}}(\bm{y}_\mathrm{T},\bm{y}_\mathrm{R})  \mathbf{s}_\mathrm{TD} + \bar{\mathbf{w}}_\mathrm{TD},
\label{eq:vectorized_TD_IO_kron}
\vspace{-1ex}
\end{equation}
where $\bar{\mathbf{H}}(\bm{y}_\mathrm{T},\bm{y}_\mathrm{R}) \in \mathbb{C}^{N d_s \times N d_s}$ explicitly reflects the dependence of the \ac{TD} transfer function of the point-to-point \ac{MIMO} system on the \ac{TX} and \ac{RX} \acp{FIM}, and is defined as
\vspace{-1ex}
\begin{equation}
\label{eq:H_bar}
\bar{\mathbf{H}}(\bm{y}_\mathrm{T},\bm{y}_\mathrm{R})\triangleq\sum_{p=1}^P (\check{\mathbf{H}}_p(\bm{y}_\mathrm{T},\bm{y}_\mathrm{R}) \otimes \mathbf{G}_p),
\vspace{-1ex}
\end{equation}
with the $N d_s$-element vectors $\mathbf{s}_\mathrm{TD}$ and $\bar{\mathbf{w}}_\mathrm{TD}$ formed by concatenating $\mathbf{s}_u$ and $\mathbf{w}_v$ from equation \eqref{eq:vectorized_TD_IO}, respectively.

For notational brevity, the matrix $\check{\mathbf{H}}_p$ in equation \eqref{eq:H_bar} will henceforth be written without explicitly noting its dependence on the \ac{TX}/\ac{RX} \ac{FIM} parameters.

\vspace{-2ex}
\subsection{OFDM Signaling}

Let $\mathcal{C}$ represent an arbitrary complex constellation set with cardinality $D$ and average energy $E_\mathrm{S}$, associated with a specific digital modulation scheme (e.g., \ac{QAM}). In \ac{OFDM}, multiple information vectors $\mathbf{x}_u \in \mathcal{C}^{N\times 1}$, where $u = \{ 1,\dots,d_s \}$, comprising a total of $Nd_s$ symbols, are modulated into the transmit signal as
\begin{equation}
\label{eq:OFDM_modulation}
\mathbf{s}^{(\text{OFDM})}_u \triangleq \mathbf{F}_N\herm  \mathbf{x}_u \in \mathbb{C}^{N \times 1},
\end{equation}
where $\mathbf{F}_N$ is the $N$-point normalized \ac{DFT} matrix. 

After circular convolution with the \ac{DD} channel and following a formulation akin to equation \eqref{eq:vectorized_TD_IO_kron}, the corresponding $Nd_s$-element discrete-time received \ac{OFDM} signal is expressed as
\begin{equation}
\label{eq:TD_OFDM_input_output}
\mathbf{r}_\text{OFDM} \triangleq \bar{\mathbf{H}}(\bm{y}_\mathrm{T},\bm{y}_\mathrm{R})  \mathbf{s}_\text{OFDM} + \bar{\mathbf{w}}_\mathrm{TD},% \in \mathbb{C}^{Nd_s \times 1},
\end{equation}
where the $Nd_s$-element vectors are defined as
\begin{equation}
\label{eq:OFDM_stacked_s}
\mathbf{s}_\text{OFDM} \triangleq 
\begin{bmatrix}
\mathbf{s}^{(\text{OFDM})}_1 \\[-1ex]
\vdots \\
\mathbf{s}^{(\text{OFDM})}_{d_s}
\end{bmatrix},\,\,
\mathbf{r}_\text{OFDM} \triangleq 
\begin{bmatrix}
\mathbf{r}^{(\text{OFDM})}_1 \\[-1ex]
\vdots \\
\mathbf{r}^{(\text{OFDM})}_{d_s}
\end{bmatrix}.
\end{equation}

At the \ac{RX}, applying \ac{OFDM} demodulation results in
\begin{equation}
\label{eq:OFDM_demodulation}
\mathbf{y}^{(\text{OFDM})}_v \triangleq \mathbf{F}_N  \mathbf{r}^{(\text{OFDM})}_v \in \mathbb{C}^{N \times 1},
\end{equation}
yielding the corresponding $Nd_s$-element discrete-time signal
\begin{equation}
\label{eq:OFDM_input_output}
\mathbf{y}_\text{OFDM} = \bar{\mathbf{H}}_\text{OFDM}(\bm{y}_\mathrm{T},\bm{y}_\mathrm{R})  \mathbf{x} + \bar{\mathbf{w}}_\text{OFDM}, 
%\in \mathbb{C}^{Nd_s \times 1},
\end{equation}

\noindent where $\bar{\mathbf{w}}_\text{OFDM} \in \mathbb{C}^{Nd_s \times 1}$ is an equivalent \ac{AWGN} with the same statistics as $\bar{\mathbf{w}}_\mathrm{TD}$, and $\bar{\mathbf{H}}_\text{OFDM}(\bm{y}_\mathrm{T},\bm{y}_\mathrm{R}) \in \mathbb{C}^{Nd_s \times Nd_s}$ denotes the effective \ac{OFDM} channel, defined similarly to $\bar{\mathbf{H}}(\bm{y}_\mathrm{T},\bm{y}_\mathrm{R})$ in equation \eqref{eq:vectorized_TD_IO_kron} as
\begin{eqnarray}
\label{eq:OFDM_effective_channel}
\bar{\mathbf{H}}_\text{OFDM}(\bm{y}_\mathrm{T},\bm{y}_\mathrm{R})
\triangleq \sum_{p=1}^P \check{\mathbf{H}}_p \otimes \overbrace{( \mathbf{F}_N \mathbf{G}_p  \mathbf{F}_N\herm)}^{\triangleq\mathbf{G}_p^\text{OFDM} \in \mathbb{C}^{N \times N}}&& \\
&&\hspace{-28.5ex}= \sum_{p=1}^P \check{\mathbf{H}}_p \otimes \mathbf{G}_p^\text{OFDM}.\nonumber
\end{eqnarray}

Note that for \ac{OFDM}, the \ac{CP} phase matrices $\mathbf{\Theta}_p$'s in equation \eqref{eq:vectorized_TD_IO} simplify to identity matrices \cite{Rou_SPM_2024}, i.e., $\phi_\mathrm{CP}(n) = 0$ in equation \eqref{eq:diagonal_CP_matrix_def}, as there is no phase offset.

% ------- CAREFUL MOVING THIS! EQUATION ON TOP OF PAGE!!! -----------
%
%
\begin{figure*}[t!]
\setcounter{equation}{26}
\normalsize
\begin{equation}
\label{eq:AFDM_diagonal_CP_matrix_def}
\bm{\varTheta}_p \triangleq \text{diag}\bigg( [ \underbrace{e^{-\jmath 2\pi {c_1} (N^2-2N\ell_p)}, e^{-\jmath 2\pi {c_1} (N^2-2N(\ell_p-1))}, \dots, e^{-\jmath 2\pi {c_1} (N^2-2N)}}_{\ell_p \; \text{terms}}, \underbrace{1, 1, \dots, 1, 1}_{N - \ell_p \; \text{ones}}] \bigg) \in \mathbb{C}^{N \times N}
\vspace{-1ex}
\end{equation}
\hrulefill
\vspace{-3ex}
\setcounter{equation}{18}
\end{figure*}
%
% ------------------------------------------------------------------
%
\subsection{OTFS Signaling}

When employing \ac{OTFS}, multiple matrices $\mathbf{X}_u \in \mathcal{C}^{\tilde{K}\times \tilde{K}'}$, with $u = \{ 1,\dots,d_s \}$, containing $\tilde{K} \tilde{K}' d_s$ symbols from an arbitrary complex constellation $\mathcal{C}$, are modulated as\footnote{For simplicity, we assume that all pulse-shaping operations utilize rectangular waveforms such that the corresponding sample matrices can be reduced to identity matrices.}
\vspace{-0.5ex}
\begin{equation}
\label{eq:TD_transmit_matrix_vectorized}
\mathbf{s}^{(\text{OTFS})}_u \triangleq \text{vec}\big(\mathbf{S}_u\big) = (\mathbf{F}_{\tilde{K}'}\herm \otimes \mathbf{I}_{\tilde{K}})  \text{vec}\big( \mathbf{X}_u
\big) \in \mathbb{C}^{\tilde{K}\tilde{K}'\times 1},
\end{equation}
where $\text{vec}(\cdot)$ denotes matrix vectorization via column stacking and $\mathbf{S}_u$ is a \ac{TD} symbols' matrix obtained from\footnote{Equivalently, $\mathbf{S}_u$ can be obtained as the Heisenberg transform of the \ac{ISFFT} of $\mathbf{X}_u$, $i.e.$, $\mathbf{S}_u = \mathbf{F}_{\tilde{K}}\herm \mathbf{X}_\text{FT}^u$ with $\mathbf{X}_\text{FT}^u \triangleq \mathbf{F}_{\tilde{K}} \mathbf{X}_u \mathbf{F}_{\tilde{K}'}\herm \in \mathbb{C}^{\tilde{K}\times \tilde{K}'}$.} the \ac{IDZT} of $\mathbf{X}_u$ as \cite{Hadani_WCNC_2017}
\begin{equation}
\label{eq:TD_transmit_matrix}
\mathbf{S}_u = \mathbf{X}_u \mathbf{F}_{\tilde{K}'}\herm  \in \mathbb{C}^{\tilde{K}\times \tilde{K}'}.
\end{equation}

It is noted that the notation in equation \eqref{eq:TD_transmit_matrix_vectorized} aligns with the approach in \cite{Raviteja_TWC_2018}, where \ac{OTFS} signals are vectorized and augmented with a \ac{CP} of length $N_\mathrm{CP}$ to mitigate inter-frame interference, similar to \ac{OFDM}.
To facilitate direct comparisons between these waveforms, we set $\tilde{K}\times \tilde{K}' = N$ henceforth.

Following transmission over the \ac{DD} channel $\bar{\mathbf{H}}(\bm{y}_\mathrm{T},\bm{y}_\mathrm{R})$ as in equation \eqref{eq:vectorized_TD_IO_kron}, the $Nd_s$-element discrete-time received \ac{OTFS} signal is modeled similarly to equation \eqref{eq:TD_OFDM_input_output} as $\mathbf{r}_\text{OTFS} \triangleq \bar{\mathbf{H}}(\bm{y}_\mathrm{T},\bm{y}_\mathrm{R})  \mathbf{s}_\text{OTFS} + \bar{\mathbf{w}}_\mathrm{TD}$, where $\mathbf{s}_\text{OTFS}$ and $\mathbf{r}_\text{OTFS}$ are defined for \ac{OTFS} analogously to equation \eqref{eq:OFDM_stacked_s}.

Unlike \ac{OFDM}, detecting the information symbols $\mathbf{X}_u$ from the $\mathbf{r}^{(\text{OTFS})}_v$ elements ($\forall v=1,\ldots,d_s$) of $\mathbf{r}_\text{OTFS}$ requires undoing the vectorization and \ac{IDZT} operations used to construct the $d_s$ elements of $\mathbf{s}_\text{OTFS}$, leading to a unique effective channel. 
Specifically, let $\bm{R}_v \triangleq \text{vec}^{-1}(\mathbf{r}^{(\text{OTFS})}_v) \in \mathbb{C}^{\tilde{K} \times \tilde{K}'}$, with $\text{vec}^{-1}(\cdot)$ indicating the de-vectorization operation according to which a vector of size $\tilde{K}\tilde{K}' \times 1$ is reshaped into a matrix of size $\tilde{K} \times \tilde{K}'$, and consider the following \ac{DZT}\footnote{Equivalently, $\mathbf{Y}_v$ can be obtained as the SFFT of the Wigner transform of $\bm{R}_v$: $\mathbf{Y}_\text{FT}^v \triangleq \mathbf{F}_{\tilde{K}} \bm{R}_v$, yielding $\mathbf{Y}_v = \mathbf{F}_{\tilde{K}}\herm \mathbf{Y}_\text{FT}^v \mathbf{F}_{\tilde{K}'}\in \mathbb{C}^{\tilde{K} \times \tilde{K}'}$.}
\begin{equation}
\label{eq:DD_rec_sig_after_SFFT}
\mathbf{Y}_v  =  \bm{R}_v \mathbf{F}_{\tilde{K}'} \in \mathbb{C}^{\tilde{K} \times \tilde{K}'}.
\end{equation}

The demodulated \ac{OTFS} signal at the \ac{RX} is then given by
\begin{equation}
\label{eq:DD_demodulation}
\mathbf{y}^{(\text{OTFS})}_v \triangleq \text{vec}(\mathbf{Y}_v) = (\mathbf{F}_{\tilde{K}'} \otimes \mathbf{I}_{\tilde{K}})  \mathbf{r}^{(\text{OTFS})}_v \in \mathbb{C}^{N\times 1},
\end{equation}
which can be compactly expressed, similar to equation \eqref{eq:OFDM_input_output}, as the following $Nd_s$-element discrete-time received signal
\begin{equation}
\label{eq:DD_input_output_relation}
\mathbf{y}_\text{OTFS} = \bar{\mathbf{H}}_\text{OTFS}(\bm{y}_\mathrm{T},\bm{y}_\mathrm{R})  \mathbf{x} + \bar{\mathbf{w}}_\text{OTFS},
%\in \mathbb{C}^{Nd_s \times 1},
\end{equation}
where $\bar{\mathbf{w}}_\text{OTFS} \in \mathbb{C}^{Nd_s \times 1}$ is an equivalent \ac{AWGN} with the same statistics as $\bar{\mathbf{w}}_\mathrm{TD}$, and $\bar{\mathbf{H}}_\text{OTFS}(\bm{y}_\mathrm{T},\bm{y}_\mathrm{R}) \in \mathbb{C}^{Nd_s \times Nd_s}$ represents the effective \ac{OTFS} channel, given by
\begin{eqnarray}
\bar{\mathbf{H}}_\text{OTFS}(\bm{y}_\mathrm{T},\bm{y}_\mathrm{R}) \triangleq \sum_{p=1}^P \check{\mathbf{H}}_p \otimes \overbrace{( (\mathbf{F}_{\tilde{K}'} \otimes \mathbf{I}_{\tilde{K}}) \mathbf{G}_p  (\mathbf{F}_{\tilde{K}'}\herm \otimes \mathbf{I}_{\tilde{K}}))}^{\triangleq\mathbf{G}_p^\text{OTFS} \in \mathbb{C}^{N \times N}} \nonumber&& \\
&&\hspace{-43ex}= \sum_{p=1}^P \check{\mathbf{H}}_p \otimes {\mathbf{G}_p^\text{OTFS}}.
\label{eq:OTFS_effective_channel}
\end{eqnarray}

Similar to \ac{OFDM}, the \ac{CP} phase matrices $\mathbf{\Theta}_p$'s reduce to identity matrices \cite{Rou_SPM_2024}. 
By comparing equations \eqref{eq:OFDM_effective_channel} and \eqref{eq:OTFS_effective_channel}, the channel modeling approach in \cite{Rou_SPM_2024} highlights both the structural similarities and the distinct effects of \ac{OFDM} and \ac{OTFS} waveforms in \ac{DD} channels.

\vspace{-1ex}
\subsection{AFDM Signaling}

The transmit signal for each information vector $\mathbf{x}_u$ using the \ac{AFDM} waveform in the considered \ac{FIM} \ac{MIMO} channel is obtained via the \ac{IDAFT} as
\begin{equation}
\label{eq:AFDM_moduation}
\mathbf{s}^{(\text{AFDM})}_u \triangleq \mathbf{\Lambda}_1\herm  \mathbf{F}_{N}\herm  \mathbf{\Lambda}_2\herm  \mathbf{x}_u \in \mathbb{C}^{N \times 1},
\end{equation}
where the $N\times N$ matrices $\mathbf{\Lambda}_i$ with $i=1,2$ are defined as
\begin{equation}
\label{eq:lambda_def}
\mathbf{\Lambda}_i \triangleq \text{diag}\big(\big[1, e^{-\jmath2\pi c_i 2^2}, \ldots, e^{-\jmath2\pi c_i (N-1)^2}\big]\big),
\end{equation}
where the first central chirp frequency $c_1$ is an optimized parameter based on maximum Doppler channel statistics \cite{Bemani_TWC_2023,Rou_SPM_2024}, and the second central chirp frequency $c_2$ is a flexible parameter that can be used for \ac{ISAC} waveform shaping \cite{Zhu_Arxiv23} or information encoding \cite{Liu_Arxiv24,RouAsilimoar2024}.

As shown in \cite{Rou_SPM_2024}, after transmission through a \ac{DD} channel, an \ac{AFDM} modulated symbol vector $\mathbf{s}^{(\text{AFDM})}_u$ with a \ac{CPP} can be modeled similarly to equation \eqref{eq:vectorized_TD_IO} by replacing the \ac{CP} matrix $\mathbf{\Theta}_p$ in equation \eqref{eq:diagonal_CP_matrix_def} with the \ac{CPP} matrix $\bm{\varTheta}_p$ given by equation \eqref{eq:AFDM_diagonal_CP_matrix_def} (top of this page). 
This requires setting the function $\phi_\mathrm{CP}(n)$ in equation \eqref{eq:diagonal_CP_matrix_def} as $\phi_\mathrm{CP}(n) = c_1 (N^2 - 2Nn)$. 
Thus, the $Nd_s$-element discrete-time received \ac{AFDM} signal can be modeled similarly to equation \eqref{eq:TD_OFDM_input_output} as $\mathbf{r}_\text{AFDM} \triangleq \bar{\mathbf{H}}(\bm{y}_\mathrm{T},\bm{y}_\mathrm{R})  \mathbf{s}_\text{AFDM} + \bar{\mathbf{w}}_\mathrm{TD}$, where $\mathbf{s}_\text{AFDM}$ and $\mathbf{r}_\text{AFDM}$ are defined for \ac{AFDM} analogously to equation \eqref{eq:OFDM_stacked_s}. 

The \ac{AFDM} demodulation of each of the $\mathbf{r}^{(\text{AFDM})}_v$ with $v\in\{1,\ldots,d_s\}$ elements of $\mathbf{r}_\text{AFDM}$ is obtained as
\setcounter{equation}{27}
\begin{equation}
\mathbf{y}^{(\text{AFDM})}_v = \mathbf{\Lambda}_2  \mathbf{F}_{N}  \mathbf{\Lambda}_1  \mathbf{r}^{(\text{AFDM})}_v \in \mathbb{C}^{N\times 1},
\label{eq:AFDM_demodulation}
\end{equation}
yielding the following expression for the $Nd_s$-element discrete-time received signal (similar to equations \eqref{eq:OFDM_input_output} and \eqref{eq:DD_input_output_relation})
\begin{equation}
\mathbf{y}_\text{AFDM} = \bar{\mathbf{H}}_\text{AFDM}(\bm{y}_\mathrm{T},\bm{y}_\mathrm{R})  \mathbf{x} + \bar{\mathbf{w}}_\text{AFDM},%\in \mathbb{C}^{Nd_s \times 1},
\label{eq:DAF_input_output_relation}
\end{equation}
where $\bar{\mathbf{w}}_\text{AFDM} \in \mathbb{C}^{Nd_s \times 1}$ is an equivalent \ac{AWGN} with the same statistics as $\bar{\mathbf{w}}_\mathrm{TD}$, and $\bar{\mathbf{H}}_\text{AFDM}(\bm{y}_\mathrm{T},\bm{y}_\mathrm{R}) \in \mathbb{C}^{Nd_s \times Nd_s}$ represents the effective \ac{AFDM} channel, given by
\begin{eqnarray}
\label{eq:AFDM_effective_channel}
\bar{\mathbf{H}}_\text{AFDM}(\bm{y}_\mathrm{T},\bm{y}_\mathrm{R}) \triangleq \sum_{p=1}^P \check{\mathbf{H}}_p \otimes \overbrace{( \mathbf{\Lambda}_2  \mathbf{F}_{N}  \mathbf{\Lambda}_1 \mathbf{G}_p \mathbf{\Lambda}_1\herm  \mathbf{F}_{N}\herm  \mathbf{\Lambda}_2\herm)}^{\mathbf{G}_p^\text{AFDM} \in \mathbb{C}^{N \times N}}&& \\
&&\hspace{-39ex}=  \sum_{p=1}^P \check{\mathbf{H}}_p \otimes {\mathbf{G}_p^\text{AFDM}}.\nonumber 
\end{eqnarray}

Evidently, equation \eqref{eq:AFDM_effective_channel} shares the same structure as equations \eqref{eq:OFDM_effective_channel} and \eqref{eq:OTFS_effective_channel}, as do the \ac{MIMO} input-output relationships in equations \eqref{eq:OFDM_input_output}, \eqref{eq:DD_input_output_relation}, and \eqref{eq:DAF_input_output_relation}. 
This suggests that signal processing techniques, such as channel estimation, can be developed within a unified framework applicable to \ac{OFDM}, \ac{OTFS}, \ac{AFDM}, and similar waveforms.

% For clarity, it is emphasized that a conventional \ac{DD}-\ac{MIMO} model—i.e., a \ac{MIMO} extension of the model in \cite{Rou_SPM_2024} without \ac{TX} and \ac{RX} \acp{FIM}—can be straightforwardly derived from the above. 
% %
% For instance, for \ac{OFDM}, \ac{OTFS}, and \ac{AFDM} waveforms, equations \eqref{eq:OFDM_effective_channel}, \eqref{eq:OTFS_effective_channel}, and \eqref{eq:AFDM_effective_channel} would yield
% %
% \begin{equation}
% \label{eq:H_DD_MIMO}
% \bar{\mathbf{H}}_\text{MIMO}\triangleq \sqrt{\tfrac{N_\mathrm{T} N_\mathrm{R}}{P}}\sum_{p=1}^P \left(h_p     \mathbf{a}_\mathrm{R}\left(\phi_p^{\rm in}\right) \mathbf{a}_\mathrm{T}\herm\left(\phi_p^{\rm out}\right)\right) \!\otimes\! {\mathbf{G}_p^\text{MIMO}}.
% \end{equation}
% %
% where the subscripts \ac{OFDM}, \ac{OTFS}, and \ac{AFDM} are replaced by the generic subscript \ac{MIMO}, and $\mathbf{a}_\mathrm{R}\left(\phi_p^{\rm in}\right)$ and $\mathbf{a}_\mathrm{T}\left(\phi_p^{\rm out}\right)$ denote typical \ac{ULA} response vectors \cite{Ranasinghe_ICASSP_2024} in a configuration with $N_\mathrm{R}$ and $N_\mathrm{T}$ receive and transmit antennas, respectively.

% % The contents added by An are marked in blue, please check
\section{FIM Optimization for ISAC}

In this paper, we aim to maximize both the communication and sensing performances by tuning both the \ac{TX}-\ac{FIM} and \ac{RX}-\ac{FIM} under the information-theoretical achievable rate criteria and a practical sensing constraint that makes sure the signal power at a certain scatterer achieves a certain threshold, as described below.

% \vspace{-8ex}
\subsection{Problem Formulation}

Under an information-theoretical approach, the \ac{FIM} optimization problem seeks to maximize the achievable rate of the \ac{MIMO} channel between a pair of \acp{FIM} by jointly optimizing the \ac{3D} surface shapes $\bm{y}_\mathrm{T}$ and $\bm{y}_\mathrm{R}$ of the transmitting and receiving \acp{FIM}, and the transmit covariance matrix $\mathbf{T} \triangleq \mathbf{x} \mathbf{x}\herm$, subject to the morphing range of these two \acp{FIM}, a total transmit power constraint $P_{\textrm{t}}$, and the sensing \ac{QoS} which aims to make sure that the signal power at the target locations achieve a certain threshold $\Psi$.
Specifically, the joint optimization problem can be formulated as
\vspace{-1ex}
\begin{subequations}\label{eq30}
\begin{alignat}{2}
&\max_{\mathbf{T},\, \bm{y}_\mathrm{T},\, \bm{y}_\mathrm{R}} &\quad& \log_{2}\det\left ( \mathbf{I}_{Nd_s}+\frac{1}{\sigma_w^{2}}\bar{\mathbf{H}}(\bm{y}_\mathrm{T},\bm{y}_\mathrm{R})\mathbf{T}\bar{\mathbf{H}}\herm(\bm{y}_\mathrm{T},\bm{y}_\mathrm{R}) \right ) \label{eq43a}\\
&\textrm{s.t.} & & \textrm{tr}\left ( \mathbf{T} \right )\leq P_{\textrm{t}},\quad \mathbf{T}\succeq \mathbf{0}, \label{eq43c}\\
& & & \textrm{tr}\left ( \bar{\mathbf{H}}(\bm{y}_\mathrm{T},\bm{y}_\mathrm{R})\mathbf{T}\bar{\mathbf{H}}\herm(\bm{y}_\mathrm{T},\bm{y}_\mathrm{R}) \right ) \geq \Psi, \label{eq_sensing}\\
& & & \bm{y}_\mathrm{T} =\left [ y_{\mathrm{T}, 1}, y_{\mathrm{T}, 2},\ldots , y_{\mathrm{T}, N_\mathrm{T}} \right ]^{T}, \label{eq43d}\\
& & & \bm{y}_\mathrm{R} =\left [ y_{\mathrm{R}, 1}, y_{\mathrm{R}, 2},\ldots , y_{\mathrm{R}, N_\mathrm{R}} \right ]^{T}, \label{eq43e}\\
& & & y_\text{min} \leq y_{\mathrm{T}, n_\mathrm{T}}\leq y_\text{max},\;\; n_\mathrm{T}=1,2,\ldots,N_\mathrm{T}, \label{eq43f}\\
& & & y_\text{min}\leq y_{\mathrm{R}, n_\mathrm{R}}\leq y_\text{max},\;\; n_\mathrm{R}=1,2,\ldots,N_\mathrm{R}, \label{eq43g}
\end{alignat}
\end{subequations}
where \eqref{eq43c} characterizes the transmit power constraint, while \eqref{eq43d} -- \eqref{eq43g} characterize the constraints on adjusting the deformation range for each antenna on the transmitting and receiving FIMs.

In addition, $\bar{\mathbf{H}}(\bm{y}_\mathrm{T},\bm{y}_\mathrm{R})$ represents the general effective channel regardless of the waveform used.
Furthermore, for clarity of exposition, let us restate that the matrices to be optimized $\check{\mathbf{H}}_p(\bm{y}_\mathrm{T},\bm{y}_\mathrm{R})  \in \mathbb{C}^{N_\mathrm{R} \times N_\mathrm{T}}$ are encapsulated inside 
\begin{equation}
\label{eq:main_def_partial}
\bar{\mathbf{H}}(\bm{y}_\mathrm{T},\bm{y}_\mathrm{R}) = \sum_{p=1}^P \Big( \check{\mathbf{H}}_p(\bm{y}_\mathrm{T},\bm{y}_\mathrm{R}) \otimes \bar{\mathbf{G}}_p \Big), %{\color{red}\triangleq (\bar{\mathbf{G}} \odot \check{\mathbf{H}} ) \mathbf{1}_P   },
\end{equation}
%
%{\color{red}with $\odot$ denoting the Khatri-Rao product, $\bar{\mathbf{G}} \triangleq [\textrm{vec}(\bar{\mathbf{G}}_1),\cdots,\textrm{vec}(\bar{\mathbf{G}}_p),\cdots,\textrm{vec}(\bar{\mathbf{G}}_P)] \in \mathbb{C}^{N^2 \times P}$ and $\check{\mathbf{H}} \triangleq [\textrm{vec}(\check{\mathbf{H}}_1),\cdots,\textrm{vec}(\check{\mathbf{H}}_p),\cdots,\textrm{vec}(\check{\mathbf{H}}_P)] \in \mathbb{C}^{N_\mathrm{R}N_\mathrm{T} \times P}$,}
%
where $\check{\mathbf{H}}_p(\bm{y}_\mathrm{T},\bm{y}_\mathrm{R})$ -- without considering digital beamforming -- can be expressed as
\begin{equation}
    \label{eq:opt_matrix_inside}
    \!\!\check{\mathbf{H}}_p(\bm{y}_\mathrm{T},\bm{y}_\mathrm{R}) \!=\! \tilde{h}_p \mathbf{b}_{\mathrm{R}:p}\left(\bm{y}_\mathrm{R},\phi_p^{\rm in},\theta_p^{\rm in}\right) \mathbf{b}_{\mathrm{T}:p}\herm\left(\bm{y}_\mathrm{T},\phi_p^{\rm out},\theta_p^{\rm out}\right),
\end{equation}
with $\bar{\mathbf{G}}_p$ representing an arbitrary channel matrix resulting from any aforementioned waveform and $\tilde{h}_p \triangleq \sqrt{\tfrac{N_\mathrm{T}N_\mathrm{R}}{P}} h_p$.

To efficiently solve this non-convex problem, we can first reformulate it leveraging a mixed objective for the regularization term by defining $g_c = \text{min}\{ \textrm{tr}\left ( \bar{\mathbf{H}}(\bm{y}_\mathrm{T},\bm{y}_\mathrm{R})\mathbf{T}\bar{\mathbf{H}}\herm(\bm{y}_\mathrm{T},\bm{y}_\mathrm{R}) \right )-\Psi , 0 \}$ as
\begin{subequations}\label{eq:reform_obj}
\begin{alignat}{2}
&\max_{\mathbf{T},\, \bm{y}_\mathrm{T},\, \bm{y}_\mathrm{R}} &\quad& \log_{2}\det\left ( \mathbf{I}_{Nd_s}+\frac{1}{\sigma_w^{2}}\bar{\mathbf{H}}(\bm{y}_\mathrm{T},\bm{y}_\mathrm{R})\mathbf{T}\bar{\mathbf{H}}\herm(\bm{y}_\mathrm{T},\bm{y}_\mathrm{R}) \right ) \nonumber\\
& & & + \beta g_c  \label{eq43a2}\\
&\textrm{s.t.} & & \textrm{tr}\left ( \mathbf{T} \right )\leq P_{\textrm{t}},\quad \mathbf{T}\succeq \mathbf{0}, \label{eq43c2}\\
& & & \bm{y}_\mathrm{T} =\left [ y_{\mathrm{T}, 1}, y_{\mathrm{T}, 2},\ldots , y_{\mathrm{T}, N_\mathrm{T}} \right ]^{T}, \label{eq43d2}\\
& & & \bm{y}_\mathrm{R} =\left [ y_{\mathrm{R}, 1}, y_{\mathrm{R}, 2},\ldots , y_{\mathrm{R}, N_\mathrm{R}} \right ]^{T}, \label{eq43e2}\\
& & & y_\text{min} \leq y_{\mathrm{T}, n_\mathrm{T}}\leq y_\text{max},\;\; n_\mathrm{T}=1,2,\ldots,N_\mathrm{T}, \label{eq43f2}\\
& & & y_\text{min}\leq y_{\mathrm{R}, n_\mathrm{R}}\leq y_\text{max},\;\; n_\mathrm{R}=1,2,\ldots,N_\mathrm{R}, \label{eq43g2}
\end{alignat}
\end{subequations}
where $\beta \in \mathbb{R}^+$ is a penalty factor denoting the tradeoff between the achievable rate and sensing \ac{QoS}.

Before solving the aforementioned optimization problem, let us clarify that the optimization of the transmit covariance matrix $\mathbf{T}$ in equation \eqref{eq:reform_obj} in the considered system\footnote{In principle, one can define a precoder $\tilde{\mathbf{U}}$ such that $\tilde{\mathbf{x}} \triangleq \tilde{\mathbf{U}} \mathbf{x}$ and $\mathbf{T} \triangleq \tilde{\mathbf{x}} \tilde{\mathbf{x}}\herm = \tilde{\mathbf{U}} \tilde{\mathbf{x}} \tilde{\mathbf{x}}\herm \tilde{\mathbf{U}}\herm$ correspondingly to optimize $\tilde{\mathbf{U}}$, but this problem is relegated to future work.} would require some non-trivial techniques such as adaptive modulation.
Since we already consider $\mathbf{x}$ to be picked from a discrete constellation such that the individual symbols are random and \ac{iid}, in the limit when the product $Nd_s$ grows large, $\mathbf{T} \rightarrow \mathbf{I}_{Nd_s}$. 
Therefore, for the sake of simplicity and to explicitly highlight the effects of the \ac{FIM}, we hereafter consider that $\mathbf{T} \approx \mathbf{I}_{Nd_s}$.
We can now obtain a suboptimal solution to the problem in equation \eqref{eq30} by solving the reformulated problem in equation \eqref{eq:reform_obj}.

\vspace{-2ex}
\subsection{Proposed Optimization Framework}
\vspace{-1ex}

Leveraging the assumption on $\mathbf{T}$, the optimization problem in equation \eqref{eq:reform_obj} can be solved for the \ac{FIM} parameters $\bm{y}_\mathrm{T}$ and $\bm{y}_\mathrm{R}$ as follows.
First, substituting $\mathbf{T} = \mathbf{I}_{Nd_s}$ yields
\vspace{-1ex}
\begin{subequations}\label{eq:reform_obj_T_fixed}
\begin{alignat}{2}
&\max_{\bm{y}_\mathrm{T},\, \bm{y}_\mathrm{R}} &\quad& \log_{2}\det\Bigg ( \mathbf{I}_{Nd_s}+\overbrace{\frac{1}{\sigma_w^{2}}\bar{\mathbf{H}}(\bm{y}_\mathrm{T},\bm{y}_\mathrm{R})\bar{\mathbf{H}}\herm(\bm{y}_\mathrm{T},\bm{y}_\mathrm{R})}^{\mathbf{Q}} \Bigg) \nonumber\\
& & & + \beta g_c  \label{eq43a3}\\
&\textrm{s.t.} & & \bm{y}_\mathrm{T} =\left [ y_{\mathrm{T}, 1}, y_{\mathrm{T}, 2},\ldots , y_{\mathrm{T}, N_\mathrm{T}} \right ]^{T}, \label{eq43d3}\\
& & & \bm{y}_\mathrm{R} =\left [ y_{\mathrm{R}, 1}, y_{\mathrm{R}, 2},\ldots , y_{\mathrm{R}, N_\mathrm{R}} \right ]^{T}, \label{eq43e3}\\
& & & y_\text{min} \leq y_{\mathrm{T}, n_\mathrm{T}}\leq y_\text{max},\;\; n_\mathrm{T}=1,2,\ldots,N_\mathrm{T}, \label{eq43f3}\\
& & & y_\text{min}\leq y_{\mathrm{R}, n_\mathrm{R}}\leq y_\text{max},\;\; n_\mathrm{R}=1,2,\ldots,N_\mathrm{R}. \label{eq43g3}
\end{alignat}
\vspace{-2ex}
\end{subequations}

However, since an optimal closed-form solution to equation \eqref{eq:reform_obj_T_fixed} is still difficult to obtain, we can leverage the gradient ascent algorithm to search for a sub-optimal solution. 
Specifically, given the surface shapes of the transmitting and receiving \acp{FIM} obtained from a previous iteration, we can adjust their surface shape configurations towards the direction of the gradient for gradually increasing the achievable rate subject to the \ac{QoS} sensing constraint as well.
The gradient ascent algorithm involves two major steps: 1)
a gradient calculation, and 2) a surface shape update.

\subsubsection{Gradient Calculation} 
In order to develop an efficient gradient-ascent algorithm, we need to first calculate the closed-form gradients of the expression $\frac{\partial \bar{\mathbf{H}}(\bm{y}_\mathrm{T},\bm{y}_\mathrm{R})}{\partial \bm{y}_\mathrm{T}}$ in a tractable manner. 
However, since $\bar{\mathbf{H}}(\bm{y}_\mathrm{T},\bm{y}_\mathrm{R})$ is a matrix and $\bm{y}_\mathrm{T}$ is a vector, the gradient $\frac{\partial \bar{\mathbf{H}}(\bm{y}_\mathrm{T},\bm{y}_\mathrm{R})}{\partial \bm{y}_\mathrm{T}}$ becomes a tensor of size $Nd_s \times Nd_s \times N_\mathrm{T}$ leading to cumbersome expressions later on.
Therefore, for ease of implementation as well as tractability, we focus on gradients per element of $\bm{y}_\mathrm{T}$, hereafter denoted by $\bm{y}_{\mathrm{T}:n_t}$, with the equivalent partial derivative denoted by $\frac{\partial \bar{\mathbf{H}}(\bm{y}_\mathrm{T},\bm{y}_\mathrm{R})}{\partial \bm{y}_{\mathrm{T}:n_t}} \in \mathbb{C}^{Nd_s \times Nd_s}$.

Let us start with the fact that $\bm{y}_{\mathrm{T}:n_t}$ only depends on $\mathbf{b}_{\mathrm{T}:p}\herm\left(\bm{y}_\mathrm{T},\phi_p^{\rm out},\theta_p^{\rm out}\right)$ to express the partial derivative of $\bar{\mathbf{H}}(\bm{y}_\mathrm{T},\bm{y}_\mathrm{R})$ with respect to $\bm{y}_{\mathrm{T}:n_t}$ using equation \eqref{eq:main_def_partial} as
\begin{equation}
    \label{eq:TX_shape_partial_derivative}
    \frac{\partial \bar{\mathbf{H}}(\bm{y}_\mathrm{T},\bm{y}_\mathrm{R})}{\partial \bm{y}_{\mathrm{T}:n_t}} = \sum_{p=1}^P \Bigg( \bigg( \tilde{h}_p \mathbf{b}_{\mathrm{R}:p}\left(\bm{y}_\mathrm{R}\right) \frac{\partial \mathbf{b}_{\mathrm{T}:p}\herm\left(\bm{y}_\mathrm{T}\right)}{\partial \bm{y}_{\mathrm{T}:n_t}} \bigg) \otimes \bar{\mathbf{G}}_p  \Bigg),
\end{equation}
where we drop the implicit dependence on the \acp{AoA} and \acp{AoD} for brevity, with $\frac{\partial \mathbf{b}_{\mathrm{T}:p}\herm\left(\bm{y}_\mathrm{T}\right)}{\partial \bm{y}_{\mathrm{T}:n_t}} \in \mathbb{C}^{1 \times N_\mathrm{T}}$ defined as
\begin{equation}
    \label{eq:explicit_gradient_TX_UPA}
    \frac{\partial \mathbf{b}_{\mathrm{T}:p}\herm\left(\bm{y}_\mathrm{T}\right)}{\partial \bm{y}_{\mathrm{T}:n_t}} \triangleq \bigg(\jmath \frac{2\pi}{\lambda} \sin(\phi_p^{\rm out}) \sin(\theta_p^{\rm out}) (\mathbf{e}_{n_t} \odot \mathbf{b}_{\mathrm{T}:p}\left(\bm{y}_\mathrm{T}\right)) \bigg)\herm,
\end{equation}
where $\mathbf{e}_{n_t} \in \mathbb{C}^{N_\mathrm{T} \times 1}$ denotes a standard basis vector with $1$ at the $n_t$-th position and zeros elsewhere while $\odot$ denotes the Hardamard product.

Next, the full gradient of $\mathbf{Q}$ with respect to $\bm{y}_{\mathrm{T}:n_t}$ -- which is a matrix of size $Nd_s \times Nd_s$ -- can then be expressed as
\begin{equation}
    \label{eq:Q_gradient}
    \frac{\partial \mathbf{Q}}{\partial \bm{y}_{\mathrm{T}:n_t}} = \frac{1}{\sigma_w^2} \bigg( \Big( \frac{\partial \bar{\mathbf{H}}}{\partial \bm{y}_{\mathrm{T}:n_t}} \Big)  \bar{\mathbf{H}}\herm + \bar{\mathbf{H}}  \Big( \frac{\partial \bar{\mathbf{H}}}{\partial \bm{y}_{\mathrm{T}:n_t}} \Big)\herm  \bigg),
\end{equation}
where we drop the inherent dependency on the variable for brevity.

Finally, denoting the objective function in equation \eqref{eq43a3} by $f \triangleq \log_{2}\det (\mathbf{I}_N + \mathbf{Q}) + \beta g_c$, the resulting scalar partial derivative per $n_t$-th element can be expressed as
\begin{equation}
    \label{eq:full_partial_f}
    \frac{\partial f}{\partial \bm{y}_{\mathrm{T}:n_t}} \!=\! \frac{1}{\ln 2} \Re \bigg\{\!\textrm{tr}\Big( (\mathbf{I}_{Nd_s} \!+\! \mathbf{Q})^{-1} \frac{\partial \mathbf{Q}}{\partial \bm{y}_{\mathrm{T}:n_t}} \Big) \!\bigg\} \!+\! \beta \frac{\partial g_c}{\partial \bm{y}_{\mathrm{T}:n_t}},
\end{equation}
where
\begin{equation}
    \label{eq:gc_gradient_TX}
    \frac{\partial g_c}{\partial \bm{y}_{\mathrm{T}:n_t}} = \Re \bigg\{ \textrm{tr}\bigg( \Big( \frac{\partial \bar{\mathbf{H}}}{\partial \bm{y}_{\mathrm{T}:n_t}} \Big)  \bar{\mathbf{H}}\herm + \bar{\mathbf{H}}  \Big( \frac{\partial \bar{\mathbf{H}}}{\partial \bm{y}_{\mathrm{T}:n_t}} \Big)\herm \bigg) \bigg\}.
\end{equation}

\begin{algorithm}[H]
\caption{FIM Optimization for ISAC}
\label{alg:proposed_decoder}
\setlength{\baselineskip}{11pt}
\textbf{Input:} Gradient descent iterations $i_{\mathrm{GD}}$ and noise variance $\sigma^2_w$. \\
\textbf{Output:} $\bm{y}_\mathrm{T}^\star$ and $\bm{y}_\mathrm{R}^\star$. 
\vspace{-1ex} 
\begin{algorithmic}[1]  
\STATEx \hspace{-3.5ex}\hrulefill
\STATEx \hspace{-3.5ex}\textbf{Initialization}
\STATEx \hspace{-3.5ex} - Set iteration counter to $i=0$.
\STATEx \hspace{-3.5ex} - Choose random $\bm{y}_\mathrm{T}^\star = \bm{y}_\mathrm{R}^\star = \bm{y}_\text{init}$ within morphing range.
\STATEx \hspace{-3.5ex} - Generate initial\footnotemark \;$\bar{\mathbf{H}}^\star(\bm{y}_\mathrm{T}^\star,\bm{y}_\mathrm{R}^\star)$ from equation \eqref{eq:main_def_partial}.
\vspace{-1ex}
\STATEx \hspace{-3.5ex}\hrulefill
\STATEx \hspace{-3.5ex}\textbf{Surface Shape Optimization}
\STATEx \hspace{-3.5ex}\textbf{for} $i=1$ to $i_{\mathrm{GD}}$ \textbf{do}:
\STATE Compute $\mathbf{Q}$ from equation \eqref{eq:reform_obj_T_fixed}.
\STATEx \textbf{for} $n_t=1$ to $N_\mathrm{T}$ \textbf{do}:
\STATE Compute $\frac{\partial \mathbf{b}_{\mathrm{T}:p}\herm\left(\bm{y}_\mathrm{T}\right)}{\partial \bm{y}_{\mathrm{T}:n_t}}, \forall p$ from equation \eqref{eq:explicit_gradient_TX_UPA}.
\STATE Compute $\frac{\partial \bar{\mathbf{H}}(\bm{y}_\mathrm{T},\bm{y}_\mathrm{R})}{\partial \bm{y}_{\mathrm{T}:n_t}}$ from equation \eqref{eq:TX_shape_partial_derivative}.
\STATE Compute $\frac{\partial \mathbf{Q}}{\partial \bm{y}_{\mathrm{T}:n_t}}$ from equation \eqref{eq:Q_gradient}.
\STATE Compute $\frac{\partial g_c}{\partial \bm{y}_{\mathrm{T}:n_t}}$ from equation \eqref{eq:gc_gradient_TX}.
\STATE Compute $\frac{\partial f}{\partial \bm{y}_{\mathrm{T}:n_t}}$ from equation \eqref{eq:full_partial_f}.
\STATEx \textbf{end for}
\STATEx \textbf{for} $n_r=1$ to $N_\mathrm{R}$ \textbf{do}:
\STATE Compute $\frac{\partial \mathbf{b}_{\mathrm{R}:p}\left(\bm{y}_\mathrm{R}\right)}{\partial \bm{y}_{\mathrm{R}:n_r}}, \forall p$ from equation \eqref{eq:explicit_gradient_RX_UPA}.
\STATE Compute $\frac{\partial \bar{\mathbf{H}}(\bm{y}_\mathrm{T},\bm{y}_\mathrm{R})}{\partial \bm{y}_{\mathrm{R}:n_r}}$ from equation \eqref{eq:RX_shape_partial_derivative}.
\STATE Compute $\frac{\partial \mathbf{Q}}{\partial \bm{y}_{\mathrm{R}:n_r}}$ from equation \eqref{eq:Q_gradient_R}.
\STATE Compute $\frac{\partial g_c}{\partial \bm{y}_{\mathrm{R}:n_r}}$ from equation \eqref{eq:gc_gradient_RX}.
\STATE Compute $\frac{\partial f}{\partial \bm{y}_{\mathrm{R}:n_r}}$ from equation \eqref{eq:full_partial_f_RX}.
\STATEx \textbf{end for}
\STATE Update the surface shapes via equation \eqref{eq:surface_shape_updtae}.

\STATEx \hspace{-3.5ex}\textbf{end for}

\end{algorithmic}
\end{algorithm}
\vspace{-2ex}

A similar procedure can now be used to calculate the partial derivative of $\bar{\mathbf{H}}(\bm{y}_\mathrm{T},\bm{y}_\mathrm{R})$ with respect to $\bm{y}_{\mathrm{R}:n_r}$ using equation \eqref{eq:main_def_partial} as
\begin{equation}
    \label{eq:RX_shape_partial_derivative}
    \frac{\partial \bar{\mathbf{H}}(\bm{y}_\mathrm{T},\bm{y}_\mathrm{R})}{\partial \bm{y}_{\mathrm{R}:n_r}} = \sum_{p=1}^P \Bigg( \bigg( \tilde{h}_p \frac{\partial \mathbf{b}_{\mathrm{R}:p}\left(\bm{y}_\mathrm{R}\right)}{\partial \bm{y}_{\mathrm{R}:n_r}} \mathbf{b}_{\mathrm{T}:p}\herm\left(\bm{y}_\mathrm{T}\right)  \bigg) \otimes \bar{\mathbf{G}}_p  \Bigg),
\end{equation}
where we once again drop the implicit dependence on the \acp{AoA} and \acp{AoD} for brevity, with
\begin{equation}
    \label{eq:explicit_gradient_RX_UPA}
    \frac{\partial \mathbf{b}_{\mathrm{R}:p}\left(\bm{y}_\mathrm{R}\right)}{\partial \bm{y}_{\mathrm{R}:n_r}} \triangleq \jmath \frac{2\pi}{\lambda} \sin(\phi_p^{\rm in}) \sin(\theta_p^{\rm in}) (\mathbf{e}_{n_t} \odot \mathbf{b}_{\mathrm{R}:p}\left(\bm{y}_\mathrm{R}\right)).
\end{equation}

Subsequently, the gradient of $\mathbf{Q}$ with respect to $\bm{y}_{\mathrm{R}:n_r}$ can then be expressed as
\begin{equation}
    \label{eq:Q_gradient_R}
    \frac{\partial \mathbf{Q}}{\partial \bm{y}_{\mathrm{R}:n_r}} = \frac{1}{\sigma_w^2} \bigg( \Big( \frac{\partial \bar{\mathbf{H}}}{\partial \bm{y}_{\mathrm{R}:n_r}} \Big)  \bar{\mathbf{H}}\herm + \bar{\mathbf{H}}  \Big( \frac{\partial \bar{\mathbf{H}}}{\partial \bm{y}_{\mathrm{R}:n_r}} \Big)\herm  \bigg),
\end{equation}
where we drop the inherent dependency on the variable for brevity.

\footnotetext{While the optimization procedure can be executed for each distinct realization of $\bar{\mathbf{H}}(\bm{y}_\mathrm{T},\bm{y}_\mathrm{R})$ which is a function of $\bar{\mathbf{G}}_p$ and other variables, in practice, after the surface shapes $\bm{y}_\mathrm{T}$ and $\bm{y}_\mathrm{R}$ have been computed for a distinct realization, the optimized surface shapes can be used irrespective to changes in delays, Doppler shifts and waveform.}

Correspondingly, the full partial derivative of the objective function with respect to $\bm{y}_{\mathrm{R}:n_r}$ can be expressed as
\begin{equation}
    \label{eq:full_partial_f_RX}
    \frac{\partial f}{\partial \bm{y}_{\mathrm{R}:n_r}} \!=\! \frac{1}{\ln 2} \Re \bigg\{\!\textrm{tr}\Big( (\mathbf{I}_{Nd_s} \!+\! \mathbf{Q})^{-1} \frac{\partial \mathbf{Q}}{\partial \bm{y}_{\mathrm{R}:n_r}} \Big) \!\bigg\} \!+\! \beta \frac{\partial g_c}{\partial \bm{y}_{\mathrm{R}:n_r}},
\end{equation}
where
\begin{equation}
    \label{eq:gc_gradient_RX}
    \frac{\partial g_c}{\partial \bm{y}_{\mathrm{R}:n_r}} = \Re \bigg\{ \textrm{tr}\bigg( \Big( \frac{\partial \bar{\mathbf{H}}}{\partial \bm{y}_{\mathrm{R}:n_r}} \Big)  \bar{\mathbf{H}}\herm + \bar{\mathbf{H}}  \Big( \frac{\partial \bar{\mathbf{H}}}{\partial \bm{y}_{\mathrm{R}:n_r}} \Big)\herm \bigg) \bigg\}.
\end{equation}

\subsubsection{Surface Shape Update}
Leveraging the gradients calculated above in equations \eqref{eq:full_partial_f} and \eqref{eq:full_partial_f_RX}, the parameter set contained in $\bm{y}_\mathrm{T}$ and $\bm{y}_\mathrm{R}$ at each $i$-th iteration can be updated as
\begin{subequations}
\label{eq:surface_shape_updtae}
    \begin{equation}
        \label{eq:TX_yT_update}
        \bm{y}_{\mathrm{T}:n_t}^{(i+1)} = \bm{y}_{\mathrm{T}:n_t}^{(i)} + \mu \frac{\partial f}{\partial \bm{y}_{\mathrm{T}:n_t}},
    \end{equation}
    \begin{equation}
        \label{eq:RX_yR_update}
        \bm{y}_{\mathrm{R}:n_r}^{(i+1)} = \bm{y}_{\mathrm{R}:n_r}^{(i)} + \mu \frac{\partial f}{\partial \bm{y}_{\mathrm{R}:n_r}},
    \end{equation}
\end{subequations}
where $\mu > 0$ represents the step size, which is typically obtained via an Armijo line search.

\section{Performance Analysis}

In this section, we analyze both the communications and sensing capabilities of the proposed system, under varying configurations and waveforms\footnote{Comparisons against \ac{RIS}-\ac{DD} and \ac{SIM}-\ac{DD} systems are relegated to future work, since this would require the formulation and solution of similar but independent achievable rate maximization problems.}.
Unless otherwise specified, the fundamental simulation parameters given in Table \ref{tab:simulation_parameters} persist throughout the section, with important parameters also given in the subtitle of the figures presented.
For the mixed term objective of equation \eqref{eq:reform_obj}, we set $\beta = 2$ for a reasonable communication-sensing tradeoff \cite{NiuWCL2024}.

\begin{table}[H]
\centering
\caption{System Parameters}
\vspace{-2ex}
\label{tab:simulation_parameters}
\begin{tabular}{|c|c|c|}
\hline
\textbf{Parameter} & \textbf{Symbol} & \textbf{Value} \\
\hline
Carrier Frequency & $f_c$ & 28 GHz \\
\hline
Carrier Wavelength & $\lambda$ & 0.0107 m \\
\hline
System Bandwidth & $B$ & 20 MHz \\
\hline
Sampling Frequency & $\lambda$ & 20 MHz \\
\hline
Number of Subcarriers & $N$ & 16, 64 \\
\hline
Total TX-FIM elements & $N_\mathrm{T}$ & 4 (2 x 2) \\
\hline
Total RX-FIM elements & $N_\mathrm{R}$ & 4 (2 x 2) \\
\hline
Total RF chains & $d_s$ & 4 \\
\hline
Number of Channel Scatterers & $P$ & 2, 5 \\
\hline
Maximum Range & $R_\text{max}$ & 120 m \\
\hline
Maximum Velocity & $V_\text{max}$ & 208 m/s \\
\hline
Maximum Morphing Range & $y_\text{max}$ & $\lambda$ m \\
\hline
Minimum Morphing Range & $y_\text{min}$ & -$\lambda$ m \\
\hline
\end{tabular}
\vspace{-1ex}
\end{table}

\vspace{-2ex}
\subsection{Communications Performance}

\begin{figure}[H]
\centering
\includegraphics[width=1\columnwidth]{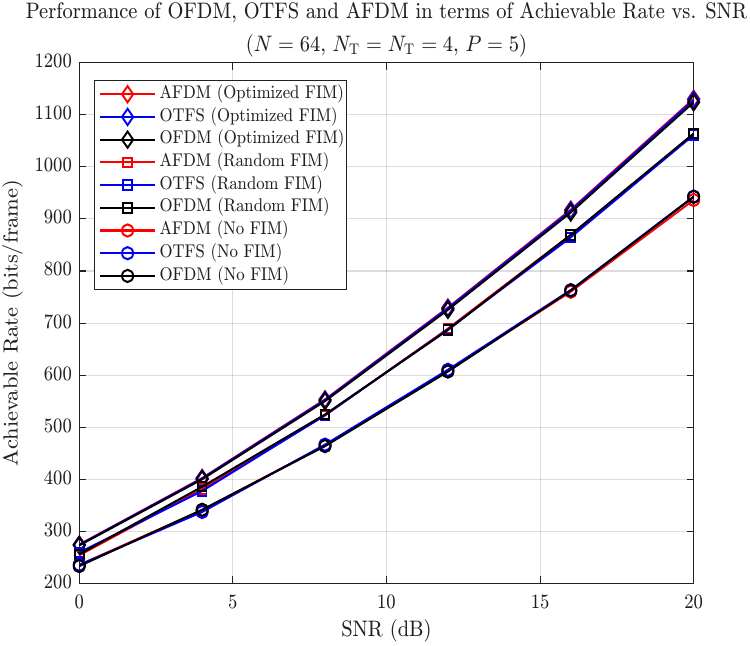}
% \vspace{-4ex}
\caption{Achievable Rate vs. \ac{SNR} of the proposed \ac{FIM} optimization strategy from Algorithm \ref{alg:proposed_decoder}.}
\label{fig:comm_N64}
\vspace{-1ex}
\end{figure}

Let us first discuss the communications performance in terms of the achievable rate as portrayed in equation \eqref{eq:reform_obj_T_fixed}.
As seen in Figure \ref{fig:comm_N64}, we present results for \ac{OFDM}, \ac{OTFS} and \ac{AFDM} under the three distinct cases with no \acp{FIM}, randomly tuned \acp{FIM} and \acp{FIM} optimized via the technique described in Algorithm \ref{alg:proposed_decoder}.
The most fundamental insight gained from these results is that having \acp{FIM} at both the \ac{TX} and the \ac{RX} significantly improve the achievable rate with a gain of around 2.5 dB going from a case with no \acp{FIM} to a case with randomly tuned \acp{FIM}.
Next, when the \acp{FIM} are optimized via Algorithm \ref{alg:proposed_decoder}, there is another gain of approximately 2 dB with respect to the case with randomly tuned \acp{FIM}.

Notably, the achievable rates for the different waveforms across all the cases are almost identical which is explained by the fact that all three schemes operate within the constraints of the same physical channel, characterized by its bandwidth, noise power, and delay-time channel impulse response. 

\vspace{-1ex}

\begin{figure}[H]
\centering
\includegraphics[width=1\columnwidth]{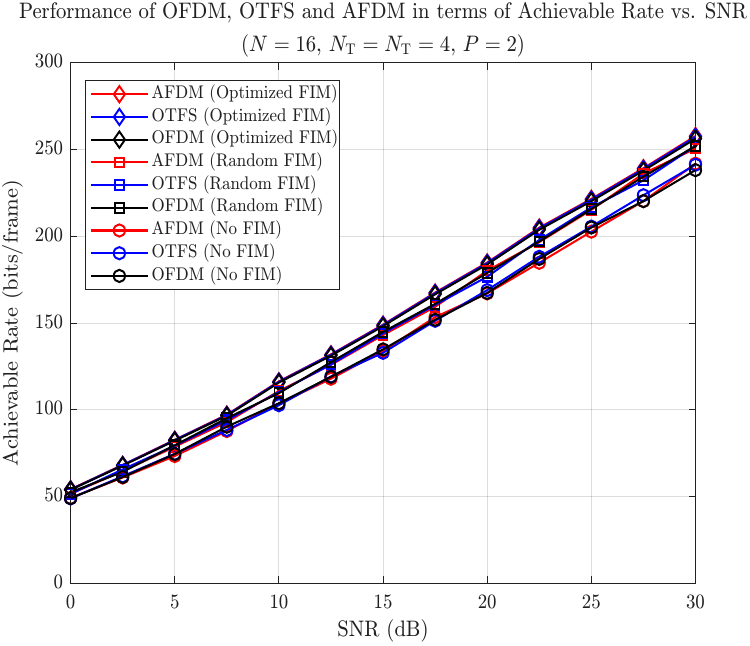}
% \vspace{-4ex}
\caption{Achievable Rate vs. \ac{SNR} of the proposed \ac{FIM} optimization strategy from Algorithm \ref{alg:proposed_decoder}.}
\label{fig:comm_N16}
\vspace{-2ex}
\end{figure}

Notably, the achievable rates for the different waveforms across all the cases are almost identical which can be explained by the fact that all three schemes operate within the constraints of the same physical channel, characterized by its bandwidth, noise power, and the fundamental delay-time channel impulse response. 
The achievable rate (or channel capacity), as defined by Shannon’s formula, is fundamentally limited by the \ac{SNR} and bandwidth, regardless of the modulation scheme.
However, this result does not translate to practical considerations such as high \ac{ICI} which significantly degrade the performance of \ac{OFDM} compared to \ac{OTFS} and \ac{AFDM} as shown in \cite{RanasingheARXIV2024}.

Next, we analyze the achievable rate for a smaller system size in Figure \ref{fig:comm_N16}.
As seen from the aforementioned figure, there is still a gain with the same trends as seen in Figure \ref{fig:comm_N64} when using the \acp{FIM}, albeit at a smaller scale due to the lower \acp{DoF} present.

\vspace{-2ex}
\subsection{Sensing Performance}
\vspace{-1ex}

After showcasing the performance gains in terms of the achievable rate for communications, let us follow through with a thorough discussion on the sensing capabilities as well.
To evaluate the sensing performance, we consider a bistatic\footnote{Notice that this setup, like in \cite{RanasingheWCNC2024}, assumes a blind setting where the \ac{RX} has no knowledge of the transmit symbols.
The procedure for a monostatic scenario is similar with the \acp{AoA} being identical to the \acp{AoD}.
However, this extension is relegated to future work since it requires the addressing of problems such as self-interference \cite{Bemani_WCL_2024}.} setup as seen in Figure \ref{fig:system_model_FIM} where the \ac{RX} aims to estimate the elevation and azimuth \acp{AoA} of all $P$ scatterers with just the knowledge of the received signals $\mathbf{y}_\text{OFDM}$, $\mathbf{y}_\text{OTFS}$ and $\mathbf{y}_\text{AFDM}$.

For this purpose, we adopt the well-known \ac{MUSIC} procedure \cite{Ranasinghe_ICASSP_2024} which is briefly summarized below\footnote{The \ac{MUSIC} spectrum can also be directly applied as a constraint in the \ac{FIM} optimization problem and this setup will be explored in future work.}.
In order to express the \ac{MUSIC} procedure in a tractable manner for all the waveforms, let us introduce the notation $\bar{\mathbf{y}}$ to denote an arbitrary received signal from any aforementioned modulation procedure.

\begin{figure}[H]
\subfigure[{\footnotesize Optimized FIM}]%
{\includegraphics[width=0.85\columnwidth]{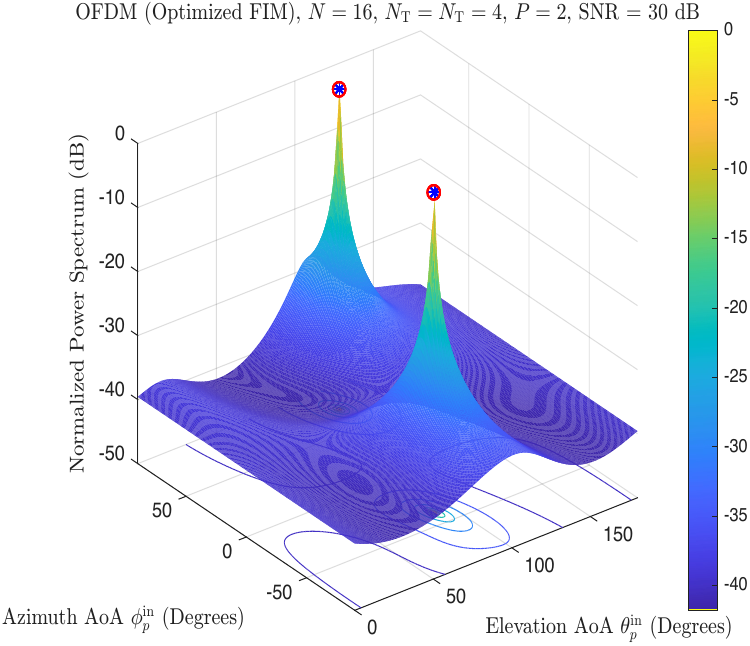}
\label{fig:Sens_OFDM_o}}\\
\subfigure[{\footnotesize Random FIM}]%
{\includegraphics[width=0.85\columnwidth]{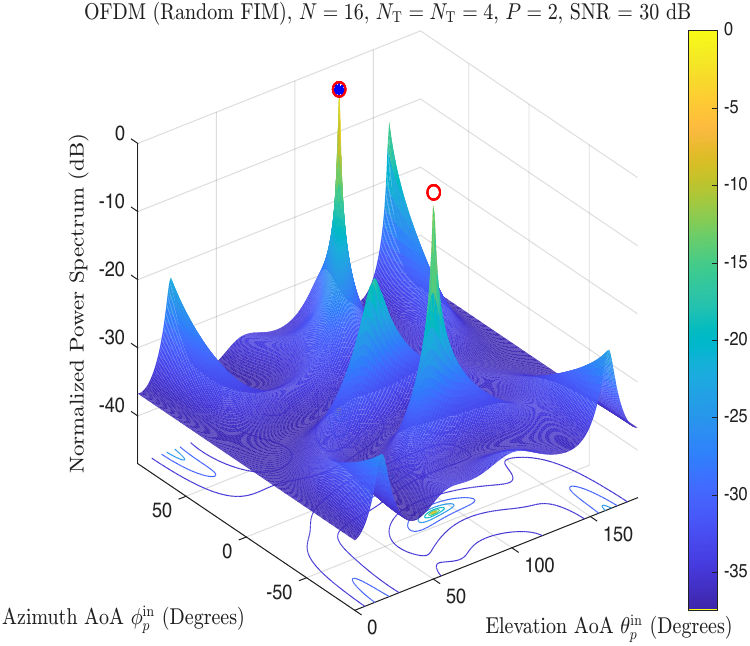}
\label{fig:Sens_OFDM_r}}
\subfigure[{\footnotesize No FIM}]%
{\includegraphics[width=0.85\columnwidth]{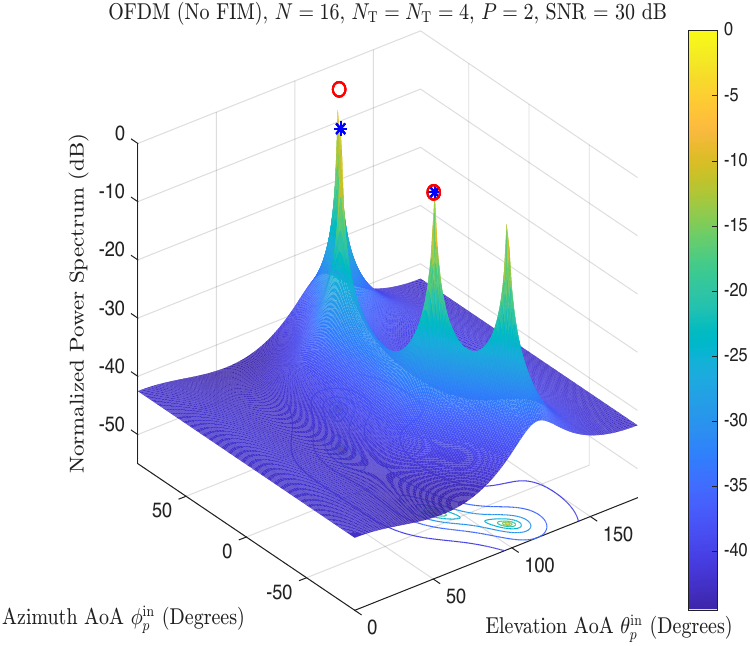}
\label{fig:Sens_OFDM_i}}
% \vspace{-2ex}
\caption{Normalized Power Spectrum for \ac{OFDM} waveforms obtained via the MUSIC algorithm at the \ac{RX} side with $P = 2$.
The red circles on each figure represents the true location of each $p$-th scatterer, while the blue stars represent the estimated positions of each $p$-th scatterer via the MUSIC algorithm.}
\label{fig:Sens_OFDM}
\vspace{-1ex}
\end{figure}

\begin{figure}[H]
\subfigure[{\footnotesize Optimized FIM}]%
{\includegraphics[width=0.85\columnwidth]{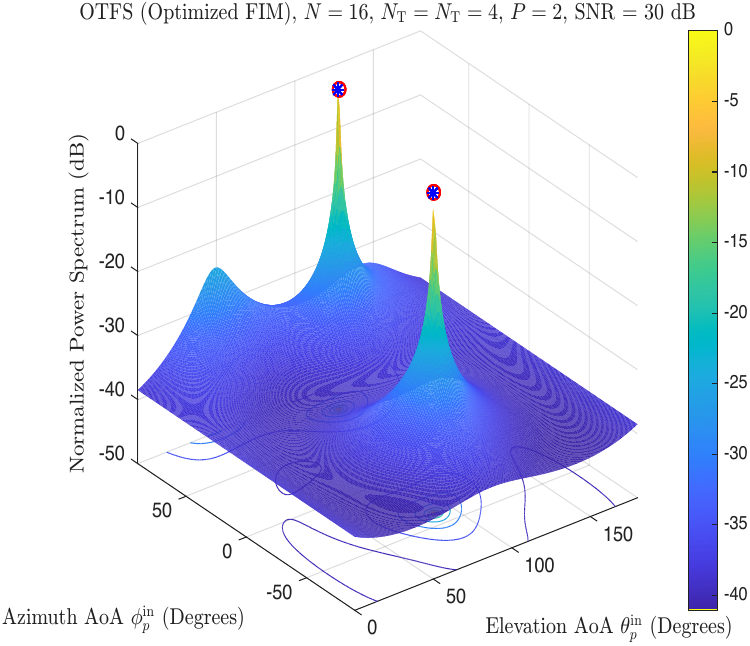}
\label{fig:Sens_OTFS_o}}\\
\subfigure[{\footnotesize Random FIM}]%
{\includegraphics[width=0.85\columnwidth]{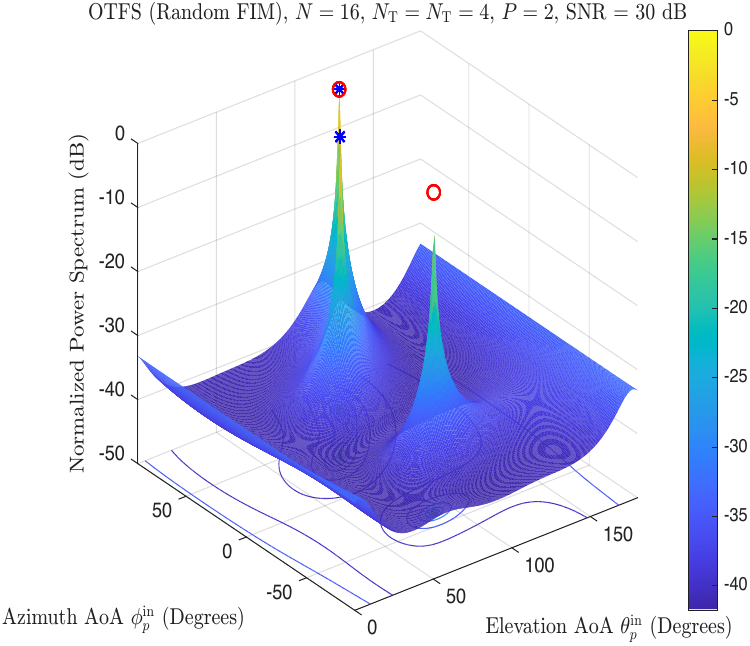}
\label{fig:Sens_OTFS_r}}
\subfigure[{\footnotesize No FIM}]%
{\includegraphics[width=0.85\columnwidth]{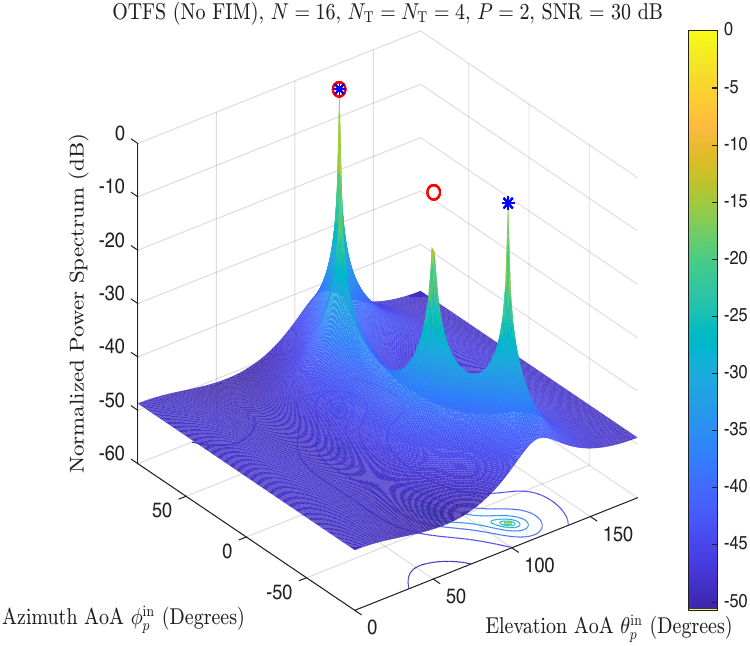}
\label{fig:Sens_OTFS_i}}
% \vspace{-2ex}
\caption{Normalized Power Spectrum for \ac{OTFS} waveforms obtained via the MUSIC algorithm at the \ac{RX} side with $P = 2$.
The red circles on each figure represents the true location of each $p$-th scatterer, while the blue stars represent the estimated positions of each $p$-th scatterer via the MUSIC algorithm.}
\label{fig:Sens_OTFS}
\vspace{-1ex}
\end{figure}

\begin{figure}[H]
\subfigure[{\footnotesize Optimized FIM}]%
{\includegraphics[width=0.85\columnwidth]{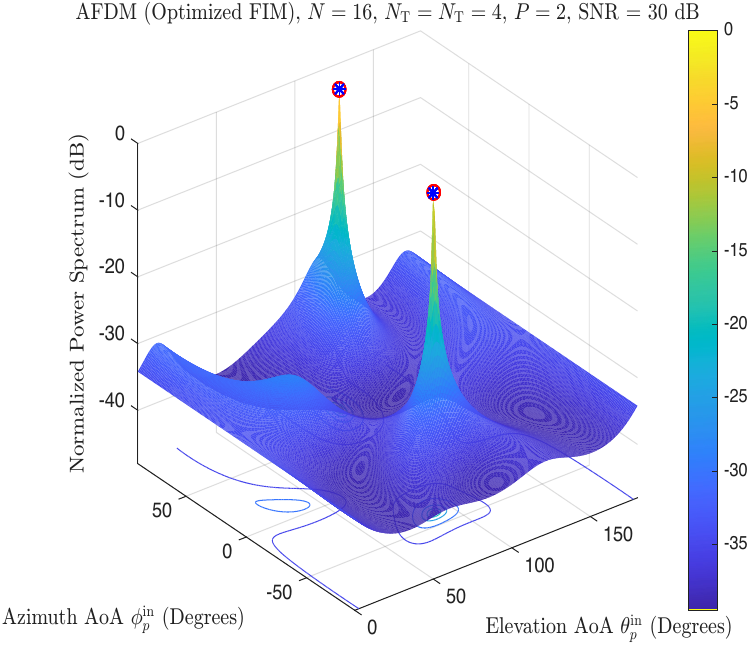}
\label{fig:Sens_AFDM_o}}\\
\subfigure[{\footnotesize Random FIM}]%
{\includegraphics[width=0.85\columnwidth]{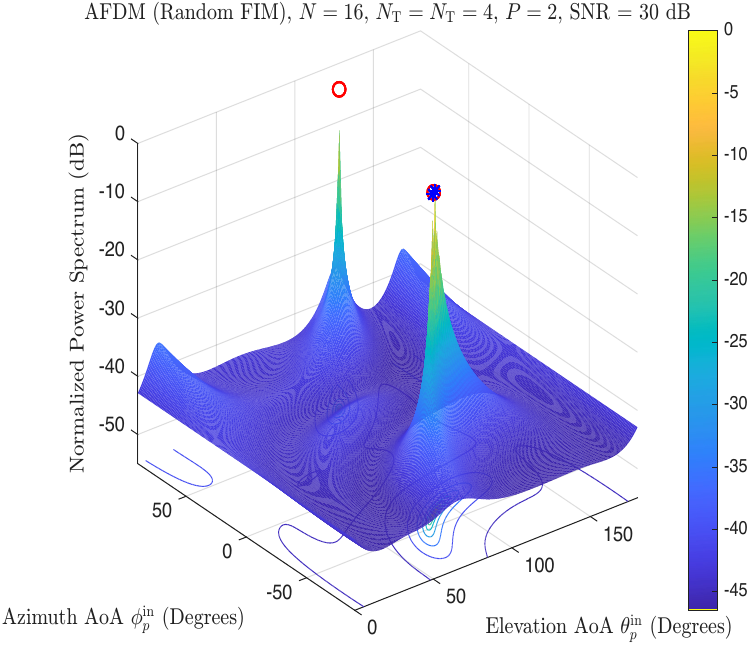}
\label{fig:Sens_AFDM_r}}
\subfigure[{\footnotesize No FIM}]%
{\includegraphics[width=0.85\columnwidth]{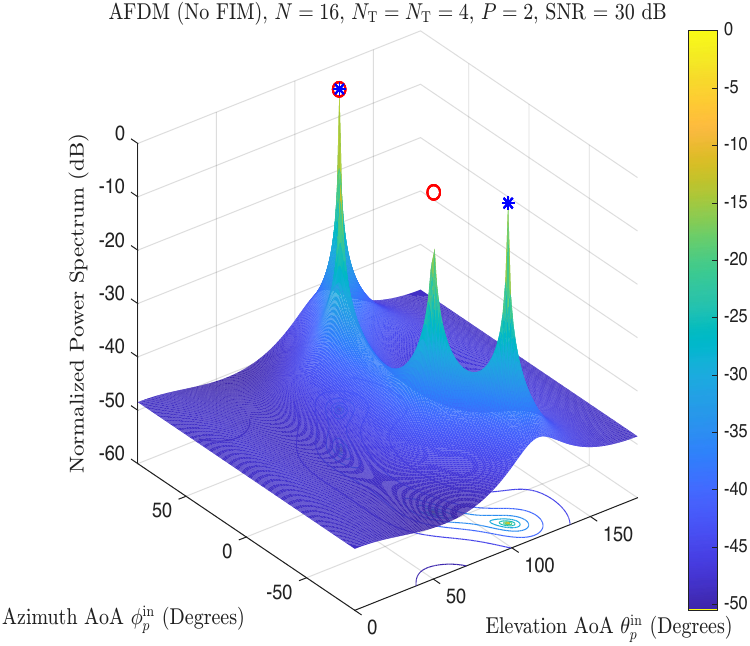}
\label{fig:Sens_AFDM_i}}
% \vspace{-2ex}
\caption{Normalized Power Spectrum for \ac{AFDM} waveforms obtained via the MUSIC algorithm at the \ac{RX} side with $P = 2$.
The red circles on each figure represents the true location of each $p$-th scatterer, while the blue stars represent the estimated positions of each $p$-th scatterer via the MUSIC algorithm.}
\label{fig:Sens_AFDM}
\vspace{-1ex}
\end{figure}

\begin{figure}[H]
\subfigure[{\footnotesize Elevation Angle Profile}]%
{\includegraphics[width=\columnwidth]{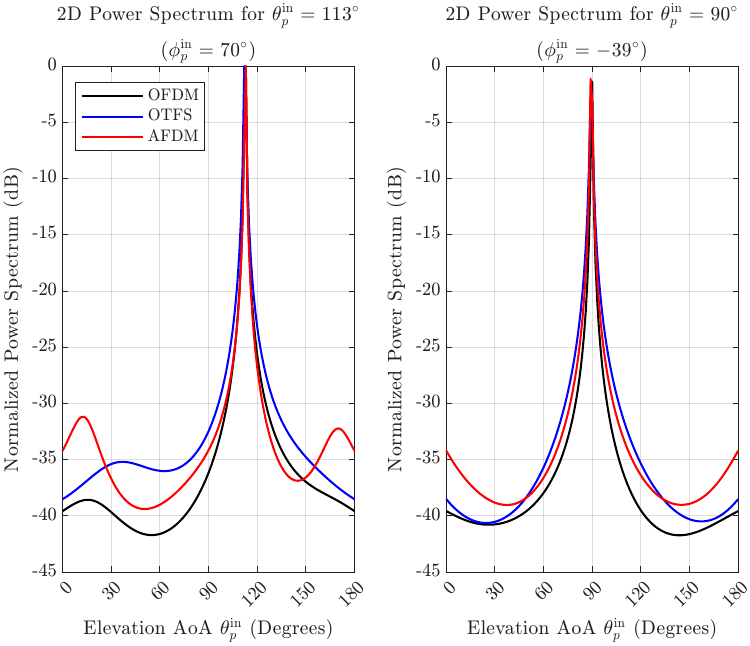}
\label{fig:Sens_2D_e}}\\
\subfigure[{\footnotesize Azimuth Angle Profile}]%
{\includegraphics[width=\columnwidth]{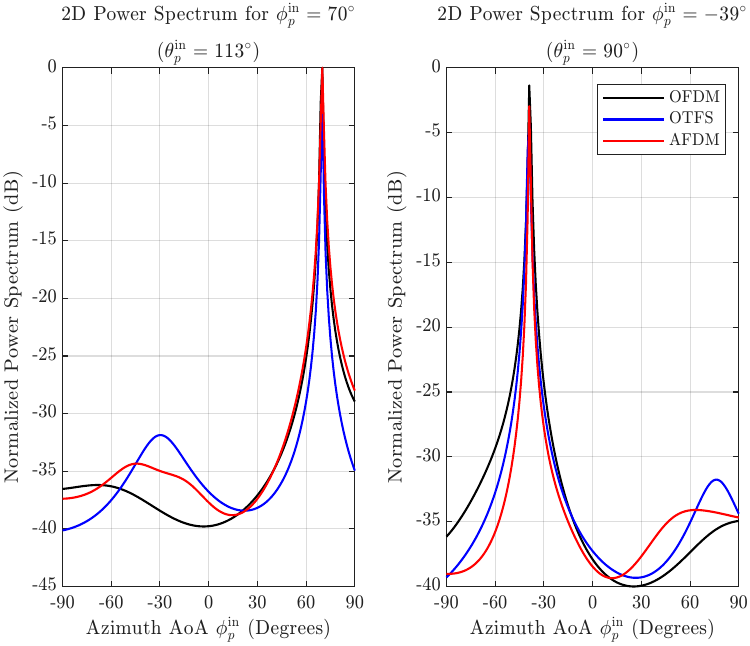}
\label{fig:Sens_2D_a}}
% \vspace{-2ex}
\caption{Normalized \ac{2D} Power Spectrum for different waveforms obtained via the MUSIC algorithm at the \ac{RX} side with $P = 2$.}
\label{fig:Sens_2D}
\vspace{-1ex}
\end{figure}

The received signal $\bar{\mathbf{y}}$ is first preprocessed via a vector inverse operation 
\begin{equation}
    \label{eq:unvec_MUSIC}
    \bar{\mathbf{Y}} = \text{vec}^{-1}(\bar{\mathbf{y}})\herm \in \mathbb{C}^{d_s \times N},
\end{equation}
where the $\text{vec}^{-1}(\cdot)$ denotes a column-wise unstacking such that the $Nd_s \times 1$ vector is shaped into a $N \times d_s$ matrix.

Next, the \ac{RX} covariance matrix $\mathbf{R}_{\bar{\mathbf{y}}}$ can be computed as
\begin{equation}
    \label{eq:RX_covariance}
    \mathbf{R}_{\bar{\mathbf{y}}} = \bar{\mathbf{Y}} \bar{\mathbf{Y}}\herm \in \mathbb{C}^{d_s \times d_s}.
\end{equation}

Denoting $\bar{\mathbf{U}} \in \mathbb{C}^{d_s \times d_s}$ to be the set of ascending eigenvectors of $\mathbf{R}_{\bar{\mathbf{y}}}$, the eigenvectors corresponding to the noise subspace of the $P$ scatterers can be extracted as $\bar{\mathbf{U}}_N \triangleq \bar{\mathbf{U}}[:,d_s-P] \in \mathbb{C}^{d_s \times P}$, which chooses the first $P$ columns of $\bar{\mathbf{U}}$.

Finally, the \ac{2D} \ac{MUSIC} spectrum for a given angle pair $(\phi_p^{\rm in},\theta_p^{\rm in})$ can be expressed as
\begin{equation}
    \label{eq:2D_MUSIC_spectrum}
    M(\phi_p^{\rm in},\theta_p^{\rm in}) = \frac{1}{\mathbf{b}_{\mathrm{R}:p}\herm\left(\bm{y}_\mathrm{R}^\star,\phi_p^{\rm in},\theta_p^{\rm in}\right) \bar{\mathbf{U}}_N \bar{\mathbf{U}}_N\herm \mathbf{b}_{\mathrm{R}:p}\left(\bm{y}_\mathrm{R}^\star,\phi_p^{\rm in},\theta_p^{\rm in}\right)},
\end{equation}
where we re-emphasize that $\mathbf{b}_{\mathrm{R}:p}\left(\bm{y}_\mathrm{R}^\star,\phi_p^{\rm in},\theta_p^{\rm in}\right)$ is the \ac{RX}-\ac{FIM} response vector computed at each pair $(\phi_p^{\rm in},\theta_p^{\rm in})$ alongside the optimal solution $\bm{y}_\mathrm{R}^\star$.

Figures \ref{fig:Sens_OFDM}, \ref{fig:Sens_OTFS} and \ref{fig:Sens_AFDM} now present results for the \ac{OFDM}, \ac{OTFS} and \ac{AFDM} waveforms respectively, in the cases where no \acp{FIM} is present, randomly tuned \acp{FIM} are present and \acp{FIM} optimized via the technique described in Algorithm \ref{alg:proposed_decoder} are present.
As seen in common across all the waveforms, in the presence of optimized \acp{FIM}, the \ac{2D} \ac{MUSIC} spectrum clearly identifies the elevation and azimuth \ac{AoA} of each scatterer with strong peaks while with randomly tuned \acp{FIM}, \ac{2D} \ac{MUSIC} yields some peaks but fails to correctly isolate the scatterers.
In addition, when there are no \acp{FIM} present, there are multiple distinct peaks that do not correspond to the actual scatterer locations and hence, the positions are not identified correctly.

Finally, to compare the performances of distinct waveforms, Figure \ref{fig:Sens_2D} offers cuts parallel to both the elevation and azimuth directions of the \ac{2D} spectra for both scatterers.
As seen from Figure \ref{fig:Sens_2D_e}, while the power of the peaks are similar across all three waveforms, \ac{OFDM} has the lowest sidelobes for the best isolation of the peaks, with \ac{OTFS} and \ac{AFDM} not far behind.
Figure \ref{fig:Sens_2D_a} also shows a similar scenario where \ac{OFDM} seems to have the lowest sidelobes albeit not by a large margin.

\vspace{-1ex}
\section{Conclusion}

We proposed a \ac{FIM}-based \ac{ISAC} \ac{DD} \ac{MIMO} channel model that proves highly effective for high-mobility scenarios.
We then formulated and solved an achievable rate maximization problem with a stringent sensing constraint using a gradient ascent algorithm, supplemented by closed-form gradients, which are shown to yield significant performance gains for the key \ac{ISAC} waveforms, namely, \ac{OFDM}, \ac{OTFS}, and \ac{AFDM}. 
Numerical results underscore the critical role of \ac{FIM} technology, highlighting that precise parametrization is essential for optimizing \ac{ISAC} performance across all evaluated waveforms, effectively mitigating the challenges of \ac{DD} channels.
An exciting possibility, to be explored in future work, is to consider the impact of \ac{FIM} on the angular resolution of \ac{AoA}/\ac{AoD} estimation, which is a key aspect of \ac{ISAC} systems.

%\bibliographystyle{IEEEtran}
%\bibliography{references}

\begin{thebibliography}{10}
\providecommand{\url}[1]{#1}
\csname url@samestyle\endcsname
\providecommand{\newblock}{\relax}
\providecommand{\bibinfo}[2]{#2}
\providecommand{\BIBentrySTDinterwordspacing}{\spaceskip=0pt\relax}
\providecommand{\BIBentryALTinterwordstretchfactor}{4}
\providecommand{\BIBentryALTinterwordspacing}{\spaceskip=\fontdimen2\font plus
\BIBentryALTinterwordstretchfactor\fontdimen3\font minus
  \fontdimen4\font\relax}
\providecommand{\BIBforeignlanguage}[2]{{%
\expandafter\ifx\csname l@#1\endcsname\relax
\typeout{** WARNING: IEEEtran.bst: No hyphenation pattern has been}%
\typeout{** loaded for the language `#1'. Using the pattern for}%
\typeout{** the default language instead.}%
\else
\language=\csname l@#1\endcsname
\fi
#2}}
\providecommand{\BIBdecl}{\relax}
\BIBdecl

\bibitem{LuongCOMMST2025}
N.~C. Luong \emph{et al.}, ``Advanced
  learning algorithms for integrated sensing and communication (ISAC) systems
  in 6G and beyond: A comprehensive survey,'' \emph{IEEE Communications Surveys
  \& Tutorials}, pp. 1--1, 2025.

\bibitem{RanasingheICNC2025_comp}
K.~R. Rayan~Ranasinghe, K.~Ando, and G.~T. Freitas~de Abreu, ``From theory to
  reality: A design framework for integrated communication and computing
  receivers,'' in \emph{2025 International Conference on Computing, Networking
  and Communications (ICNC)}, 2025, pp. 865--870.

\bibitem{ranasinghe2025flexibledesignframeworkintegrated}
\BIBentryALTinterwordspacing
K.~R.~R. Ranasinghe, K.~Ando, H.~S. Rou, G.~T.~F. de~Abreu, T.~Takahashi, M.~D.
  Renzo, and D.~G. Gonzalez, ``A flexible design framework for integrated
  communication and computing receivers,'' 2025. [Online]. Available:
  \url{https://arxiv.org/abs/2506.05944}
\BIBentrySTDinterwordspacing

\bibitem{YuNWTWORK2023}
H.~Yu, M.~Shokrnezhad, T.~Taleb, R.~Li, and J.~Song, ``Toward 6G-based
  metaverse: Supporting highly-dynamic deterministic multi-user extended
  reality services,'' \emph{IEEE Network}, vol.~37, no.~4, pp. 30--38, 2023.

\bibitem{NguyenJSAC2024}
V.-D. Nguyen \emph{et al.}, ``Network-aided intelligent
  traffic steering in 6G O-Ran: A multi-layer optimization framework,''
  \emph{IEEE Journal on Selected Areas in Communications}, vol.~42, no.~2, pp.
  389--405, 2024.

\bibitem{CuiCC2022}
H.~Cui \emph{et al.}, ``Space-air-ground integrated network (SAGIN) for 6G: Requirements,
  architecture and challenges,'' \emph{China Communications}, vol.~19, no.~2,
  pp. 90--108, 2022.

\bibitem{DengJSAC2023}
R.~Deng \emph{et al.},
  ``Reconfigurable holographic surfaces for ultra-massive MIMO in 6G: Practical
  design, optimization and implementation,'' \emph{IEEE Journal on Selected
  Areas in Comm.}, vol.~41, no.~8, 2023.

\bibitem{Chowdhury_6G}
M.~Z. Chowdhury \emph{et al.}, ``6G wireless
  communication systems: Applications, requirements, technologies, challenges,
  and research directions,'' \emph{IEEE Open Journal of the Communications
  Society}, vol.~1, 2020.

\bibitem{Dang_6G}
S.~Dang, O.~Amin, B.~Shihada, and M.-S. Alouini, ``What should 6G be?''
  \emph{Nature Electronics}, vol.~3, no.~1, pp. 20--29, 2020.

\bibitem{GiordaniCOMMAG2020}
M.~Giordani, M.~Polese, M.~Mezzavilla, S.~Rangan, and M.~Zorzi, ``Toward 6G networks: Use cases and technologies,'' \emph{IEEE Communications Magazine},
  vol.~58, no.~3, pp. 55--61, 2020.

\bibitem{Bliss_Govindasamy_2013}
D.~W. Bliss and S.~Govindasamy.\hskip 1em plus 0.5em minus 0.4em\relax
  Cambridge University Press, 2013.

\bibitem{TariqWCOMM2020}
F.~Tariq \emph{et al.}, ``A speculative study on 6G,'' \emph{IEEE Wireless
  Communications}, vol.~27, no.~4, pp. 118--125, 2020.

\bibitem{Rou_SPM_2024}
H.~S. Rou \emph{et al.}, ``From orthogonal time–frequency space to affine
  frequency-division multiplexing: A comparative study of next-generation
  waveforms for integrated sensing and communications in doubly dispersive
  channels,'' \emph{IEEE Signal Process. Mag.}, vol.~41, no.~5, 2024.

\bibitem{Hadani_WCNC_2017}
R.~Hadani \emph{et al.}, ``Orthogonal time frequency space modulation,'' in \emph{Proc.
  IEEE WCNC}, San Francisco, USA, 2017.

\bibitem{liu2022integrated}
F.~Liu \emph{et al.},
  ``Integrated sensing and communications: Toward dual-functional wireless
  networks for 6G and beyond,'' \emph{IEEE Journal on Selected Areas in
  Comm.}, vol.~40, no.~6, pp. 1728--1767, 2022.

\bibitem{NguyenJSTSP2024}
N.~T. Nguyen \emph{et al.}, ``Joint communications and sensing hybrid beamforming design via
  deep unfolding,'' \emph{IEEE Journal of Selected Topics in Signal
  Processing}, vol.~18, no.~5, pp. 901--916, 2024.

\bibitem{XiaoTSP2024}
Z.~Xiao \emph{et al.}, ``A novel joint
  angle-range-velocity estimation method for MIMO-OFDM ISAC systems,''
  \emph{IEEE Transactions on Signal Processing}, vol.~72, pp. 3805--3818, 2024.

\bibitem{RanasingheARXIV2024}
K.~R.~R. Ranasinghe \emph{et al.}, ``Joint channel, data, and radar parameter estimation for AFDM systems in doubly-dispersive channels,'' \emph{IEEE Transactions on Wireless
  Communications}, vol.~24, no.~2, 2025.

\bibitem{LuoIoTJ2025}
Y.~Luo \emph{et al.}, ``A novel
  angle-delay-doppler estimation scheme for AFDM-ISAC system in mixed
  near-field and far-field scenarios,'' \emph{IEEE Internet of Things Journal},
  vol.~12, no.~13, 2025.

\bibitem{cheng2022integrated}
X.~Cheng, D.~Duan, S.~Gao, and L.~Yang, ``Integrated sensing and communications
  (ISAC) for vehicular communication networks (VCN),'' \emph{IEEE Internet of
  Things Journal}, vol.~9, no.~23, pp. 23\,441--23\,451, 2022.

\bibitem{AlexandropolousVTM2024}
G.~C. Alexandropoulos \emph{et al.}, ``Hybrid reconfigurable intelligent metasurfaces: Enabling
  simultaneous tunable reflections and sensing for 6G wireless
  communications,'' \emph{IEEE Vehicular Technology Magazine}, vol.~19, no.~1,
  pp. 75--84, 2024.

\bibitem{StutzOJCOMS2025}
A.~Stutz-Tirri \emph{et al.}, ``Efficient and physically consistent
  modeling of reconfigurable electromagnetic structures,'' \emph{IEEE Open
  Journal of the Communications Society}, vol.~6, pp. 1610--1633, 2025.

\bibitem{dardari2025overtheairmultifunctionalwidebandelectromagnetic}
\BIBentryALTinterwordspacing
D.~Dardari, ``Over-the-air multifunctional wideband electromagnetic signal
  processing using dynamic scattering arrays,'' 2025. [Online]. Available:
  \url{https://arxiv.org/abs/2506.00619}
\BIBentrySTDinterwordspacing

\bibitem{AtaloglouTAP2025}
V.~G. Ataloglou and G.~V. Eleftheriades, ``A reconfigurable intelligent surface
  with surface-wave assisted beamforming capabilities,'' \emph{IEEE
  Transactions on Antennas and Propagation}, pp. 1--1, 2025.

\bibitem{LSA_2014_Cui_Coding}
T.~J. Cui \emph{et al.}, ``Coding metamaterials,
  digital metamaterials and programmable metamaterials,'' \emph{Light Sci. \& Appl.}, vol.~3, no.~10, 2014.

\bibitem{KolomvakisTWC2025}
N.~Kolomvakis, A.~Kosasih, and E.~Bj{\"o}rnson, ``Nonlinear distortion radiated
  from large arrays and active reconfigurable intelligent surfaces,''
  \emph{IEEE Transactions on Wireless Communications}, vol.~24, no.~6, 2025.

\bibitem{DrouliasTWC2024}
S.~Droulias, G.~Stratidakis, E.~Bj{\"o}rnson, and A.~Alexiou, ``Reconfigurable
  intelligent surfaces as spatial filters,'' \emph{IEEE Transactions on
  Wireless Communications}, vol.~23, no.~11, pp. 16\,922--16\,934, 2024.

\bibitem{ZhangTWC2025}
X.~Zhang, H.~Zhang, L.~Liu, Z.~Han, H.~V. Poor, and B.~Di, ``Target detection
  and positioning aided by reconfigurable surfaces: Reflective or
  holographic?'' \emph{IEEE Transactions on Wireless Communications}, vol.~23,
  no.~12, pp. 19\,215--19\,230, 2024.

\bibitem{AnWC2024}
J.~An \emph{et al.}, ``Stacked intelligent metasurface-aided {MIMO} transceiver
  design,'' \emph{IEEE Wireless Commun.}, vol.~31, no.~4, pp. 123--131, 2024.

\bibitem{AXN2023}
J.~An \emph{et al.}, ``Stacked intelligent metasurfaces for efficient holographic {MIMO}  communications in {6G},'' \emph{IEEE J. Sel. Areas Commun.}, vol.~41, no.~8,
  pp. 2380--2396, 2023.

\bibitem{TWC_2024_An_Flexible}
J.~An \emph{et al.}, ``Flexible  intelligent metasurfaces for downlink multiuser {MISO} communications,''
  \emph{IEEE Trans. Wireless Commun.}, vol.~24, no.~4, pp. 2940--2955, 2025.

\bibitem{Nature_2022_Bai_A}
Y.~Bai \emph{et~al.}, ``A dynamically reprogrammable surface with
  self-evolving shape morphing,'' \emph{Nature}, vol. 609, no. 7928, pp.
  701--708, Sep. 2022.

\bibitem{TAP_2025_An_Emerging}
J.~An, M.~Debbah, T.~J. Cui, Z.~N. Chen, and C.~Yuen, ``Emerging technologies
  in intelligent metasurfaces: Shaping the future of wireless communications,''
  \emph{IEEE Trans. Antennas Propag.}, pp. 1--1, 2025.

\bibitem{Sci_2015_Ni_An}
X.~Ni \emph{et al.}, ``An ultrathin invisibility skin cloak for visible light,'' \emph{Sci.}, vol. 349, no. 6254, pp. 1310--1314, Sep. 2015.

\bibitem{NC_2016_Kamali_Decoupling}
S.~M. Kamali \emph{et al.}, ``Decoupling optical function and geometrical form using conformal flexible dielectric  metasurfaces,'' \emph{Nature Commun.}, vol.~7, no.~1, p. 11618, May 2016.

\bibitem{NC_2022_Ni_Soft}
X.~Ni \emph{et~al.}, ``Soft shape-programmable surfaces by fast
  electromagnetic actuation of liquid metal networks,'' \emph{Nature Commun.},
  vol.~13, no.~1, p. 5576, Sep. 2022.

\bibitem{TCOM_2025_An_Flexible}
J.~An \emph{et al.}, ``Flexible
  intelligent metasurfaces for enhancing {MIMO} communications,'' \emph{IEEE
  Trans. Commun.}, pp. 1--15, 2025, Early Access.

\bibitem{TVT_2025_Teng_Flexible}
Z.~Teng \emph{et al.}, ``Flexible intelligent
  metasurface for enhancing multi-target wireless sensing,'' \emph{IEEE Trans.
  Veh. Technol.}, pp. 1--6, 2025.

\bibitem{ranasinghe2025metasurfacesintegrateddoublydispersivemimochannel}
\BIBentryALTinterwordspacing
K.~R.~R. Ranasinghe \emph{et al.}, ``Metasurfaces-integrated doubly-dispersive MIMO: Channel
  modeling and optimization,'' 2025. [Online]. Available:
  \url{https://arxiv.org/abs/2506.14985}
\BIBentrySTDinterwordspacing

\bibitem{ranasinghe2025doublydispersivemimochannelsstacked}
\BIBentryALTinterwordspacing
K.~R.~R. Ranasinghe \emph{et al.}, ``Doubly-dispersive MIMO channels with stacked intelligent
  metasurfaces: Modeling, parametrization, and receiver design,'' 2025.
  [Online]. Available: \url{https://arxiv.org/abs/2501.07724}
\BIBentrySTDinterwordspacing

\bibitem{ranasinghe2025parametrizedstackedintelligentmetasurfaces}
\BIBentryALTinterwordspacing
K.~R.~R. Ranasinghe, I.~A.~M. Sandoval, G.~T.~F. de~Abreu, and G.~C.
  Alexandropoulos, ``Parametrized stacked intelligent metasurfaces for bistatic
  integrated sensing and communications,'' 2025. [Online]. Available:
  \url{https://arxiv.org/abs/2504.20661}
\BIBentrySTDinterwordspacing

\bibitem{AnJSAC2023}
J.~An \emph{et al.}, ``Stacked intelligent metasurfaces for efficient holographic {MIMO}
  communications in {6G},'' \emph{IEEE J. Sel. Areas Commun.}, vol.~41, no.~8,
  pp. 2380--2396, 2023.

\bibitem{AnTWC2025}
J.~An, C.~Yuen \emph{et al.},
  ``Flexible intelligent metasurfaces for downlink multiuser miso
  communications,'' \emph{IEEE Transactions on Wireless Communications}, pp.
  1--1, 2025.

\bibitem{AnJSAC2024}
J.~An \emph{et al.},
  ``Two-dimensional direction-of-arrival estimation using stacked intelligent
  metasurfaces,'' \emph{IEEE J. Sel. Areas Commun.}, vol.~42, no.~10, pp.
  2786--2802, 2024.

\bibitem{SrivastavaTWC2022}
S.~Srivastava, R.~K. Singh, A.~K. Jagannatham, A.~Chockalingam, and L.~Hanzo,
  ``Otfs transceiver design and sparse doubly-selective csi estimation in
  analog and hybrid beamforming aided mmwave mimo systems,'' \emph{IEEE
  Transactions on Wireless Communications}, vol.~21, no.~12, pp.
  10\,902--10\,917, 2022.

\bibitem{YanCommL2023}
Y.~Yan, C.~Shan, J.~Zhang, and H.~Zhao, ``Off-grid channel estimation for
  {OTFS}-based mmwave hybrid beamforming systems,'' \emph{IEEE Commun. Lett.},
  vol.~27, no.~8, pp. 2167--2171, 2023.

\bibitem{RanasingheICNC2025_oversampling}
K.~R. Rayan~Ranasinghe, Y.~Ge, G.~T. Freitas~de Abreu, and Y.~Liang~Guan, ``Joint channel estimation and data detection for AFDM receivers with oversampling,'' in \emph{Proc. International Conference on Computing, Networking and Communications (ICNC)}, 2025.

\bibitem{Raviteja_TWC_2018}
P.~Raviteja \emph{et al.}, ``Interference cancellation and iterative detection for orthogonal time frequency space modulation,'' \emph{IEEE Trans. Wireless Commun.}, vol.~17, no.~10, 2018.

\bibitem{Bemani_TWC_2023}
A.~Bemani, N.~Ksairi, and M.~Kountouris, ``Affine frequency division
  multiplexing for next generation wireless communications,'' \emph{IEEE Trans.
  Wireless Commun.}, vol.~22, no.~11, pp. 8214--8229, 2023.

\bibitem{Zhu_Arxiv23}
J.~Zhu, Y.~Tang, X.~Wei, H.~Yin, J.~Du, Z.~Wang, and Y.~Liu, ``A low-complexity
  radar system based on affine frequency division multiplexing modulation,''
  \emph{arXiv preprint arXiv:2312.11125}, 2023.

\bibitem{Liu_Arxiv24}
G.~Liu, T.~Mao, R.~Liu, and Z.~Xiao, ``Pre-chirp-domain index modulation for affine frequency division multiplexing,'' \emph{arXiv preprint
  arXiv:2402.15185}, 2024.

\bibitem{RouAsilimoar2024}
\BIBentryALTinterwordspacing
H.~S. Rou \emph{et al.},
  ``AFDM chirp-permutation-index modulation with quantum-accelerated codebook design,'' IEEE Proceedings of the Asilomar Conference on Signals, Systems and Computers (ASILOMAR) 2024. [Online]. Available:
  \url{https://arxiv.org/abs/2405.02085}
\BIBentrySTDinterwordspacing

\bibitem{Ranasinghe_ICASSP_2024}
K.~R.~R. Ranasinghe, H.~S. Rou, and G.~T.~F. de~Abreu, ``Fast and efficient
  sequential radar parameter estimation in {MIMO}-{OTFS} systems,'' in
  \emph{Proc. IEEE ICASSP}, Seoul, South Korea, 2024.

\bibitem{NiuWCL2024}
H.~Niu, J.~An, A.~Papazafeiropoulos, L.~Gan, S.~Chatzinotas, and M.~Debbah,
  ``Stacked intelligent metasurfaces for integrated sensing and
  communications,'' \emph{IEEE Wireless Communications Letters}, 2024.

\bibitem{RanasingheWCNC2024}
K.~R.~R. Ranasinghe \emph{et al.}, ``Blind bistatic radar parameter estimation for AFDM systems in doubly-dispersive channels,'' in \emph{Proc. IEEE Wireless Communications and Networking Conference (WCNC)}, 2024. [Online]. Available: \url{https://arxiv.org/abs/2407.05328}


\bibitem{Bemani_WCL_2024}
A.~Bemani, N.~Ksairi, and M.~Kountouris, ``Integrated sensing and
  communications with affine frequency division multiplexing,'' \emph{IEEE Wireless Commun. Lett.}, early access, 2024.
\end{thebibliography}


% \newpage

\end{document}